\documentclass[a4paper,11pt]{article}
\pdfoutput=1 
\usepackage{jcappub} 
\usepackage{verbatim} 
\usepackage[T1]{fontenc} 
\usepackage[normalem]{ulem}

\title{\fontsize{18}{25}\selectfont\boldmath  The Goldstone Awakens: \\ Unimodular dark energy in scale-invariant $R^2$ gravity}

\author[a,1]{Mariaveronica De Angelis\note{Corresponding author.}}
\author[a,b]{Javier Rubio}
\affiliation[a]{Departamento de Física Teórica, Facultad de Ciencias Físicas, Universidad Complutense de Madrid, 28040 Madrid, Spain}
\affiliation[b]{Instituto de Física de Partículas y del Cosmos (IPARCOS), Universidad Complutense de Madrid, 28040 Madrid, Spain}
\emailAdd{mdeangel@ucm.es}
\emailAdd{javier.rubio@ucm.es}

\abstract{
We construct an $R^2$ cosmology based on scale invariance and unimodular gravity in which the Goldstone boson of dilatations establishes a predictive connection between inflation and dark energy. The conservation of the corresponding Noether current confines the inflationary trajectory to a one-dimensional orbit in field space, freezing the Goldstone direction and rendering the inflationary dynamics effectively single field. After inflation, the system settles on a Minkowski vacuum manifold. The unimodular integration constant, already present in the full theory but negligible during inflation, lifts the flat Goldstone direction, generating an exponential potential for the canonically normalised field. Unlike in phenomenological quintessence models, its slope is not a free DE parameter, but is instead fixed by the field-space geometry inherited from the inflationary attractor. This geometry excludes matter-era tracking and, throughout the inflationary viability region, places the field in the thawing regime compatible with accelerated expansion. For a representative inflationary benchmark, the now pseudo-Goldstone field remains frozen until close to the present epoch, yielding DE equation-of-state parameters $(w_0,w_a)\simeq(-0.992,-0.011)$ in the CPL parametrisation $w(a)=w_0+w_a(1-a)$. More generally, the same parameter controlling the inflationary spectral tilt also fixes the slope of the DE potential, leading to a tight correlation between early- and late-Universe observables: increasing the thawing signal lowers $n_s$, while values of $n_s$ closer to unity drive the model towards $\Lambda$CDM. 
}

\begin{document}
\maketitle
\flushbottom

\section{Introduction}\label{sec:introduction}

The absence of any fundamental mass scale in the classical action is a powerful organising principle for theories of gravity and matter. Scale-invariant theories in which every operator carries mass dimension four may generate all dimensionful quantities exclusively through spontaneous symmetry breaking or as integration constants of the equations of motion ~\cite{Fujii:1974,Englert:1976,tHooft:1980,Wetterich:1988,Bardeen:1995kv,Meissner:2007,Einhorn:2014gfa}. Related constructions use approximate conformal symmetry to connect the electroweak and gravitational scales and to formulate inflation without introducing fundamental mass parameters \cite{Shaposhnikov:2018xkv,Shaposhnikov:2018jag,Shaposhnikov:2020geh,Karananas:2020qkp}.  In cosmology, this has two immediate and non-trivial consequences. First, the inflationary potential is protected against additive mass renormalisation, so the flatness required for slow-roll inflation follows from the symmetry rather than from a fine-tuning of the action~\cite{Wetterich:1988,Bezrukov:2008,Shaposhnikov:2010,Salvio:2014,Bezrukov:2017dyv}. Second, classical scale invariance forbids an explicit field-independent vacuum-energy term in the Jordan-frame Lagrangian; it can only emerge as an integration constant of the equations of motion. It is this second observation that makes the combination of scale invariance (SI) and unimodular gravity (UG) particularly natural and provides the central motivation for the present work. 

Scale-invariant theories of gravity and scalar fields have been studied extensively as models of inflation~\cite{    Blas:2011ac,Salvio:2014,Karananas:2016kyt,Rubio:2017gty,Tambalo:2017,Ferreira:2018,Casas:2018fum,Rubio:2019,Ferreira:2019,Ghilencea:2019,Karam:2018mft,Shaposhnikov:2020frq,Ghilencea:2020,Bettoni:2021qfs,vandeBruck:2021xkm,Rinaldi:2023mdf,Cecchini:2024,Karananas:2025fas}. One of the simplest realisations in four dimensions contains an $R^2$ term leading to a scalaron degree of freedom, a scalar non-minimally coupled through $\phi^2R$, and a quartic self-interaction $\phi^4$, with no mass terms permitted by the symmetry \cite{Ferreira:2019,Rinaldi:2015,Rinaldi:2023mdf,Cecchini:2024}. Inflation in this class of models is generically of the Starobinsky type, with the tensor-to-scalar ratio $r$ predicted to lie in the range $0.0026<r<0.0033$, and therefore within the reach of next-generation CMB polarisation experiments~\cite{CMBS4:2016,LiteBIRD:2023,SimonsObs:2019}. A central structural property of these theories is that the Noether current associated with SI constrains the classical trajectory to a one-dimensional orbit in field space~\cite{Garcia-Bellido:2011kqb,Ferreira:2018}. One of the two scalar degrees of freedom---the Goldstone boson of spontaneously broken scale symmetry---is frozen during inflation, while the remaining scalaron drives an effective single-field slow-roll evolution. Moreover, isocurvature perturbations do not source the curvature mode \cite{Garcia-Bellido:2011kqb,Ferreira:2018}. For the specific quadratic-gravity model considered here, this picture was confirmed by a full numerical integration of the two-field equations of motion in~\cite{Rinaldi:2023mdf,Cecchini:2024}. In the present work, we focus on the Minkowski branch on which the inflationary potential vanishes at its minimum. With this choice, the unimodular lifting becomes solely responsible for the late-time dark-energy (DE) component.

UG is a minimal modification of general relativity in which the metric determinant is constrained to a fixed value~\cite{Anderson:1971,Buchmuller:1988a,Unruh:1989,Henneaux:1989,VanDerBij:1982,Percacci:2018,Ellis:2011,Alvarez:2005,Carballo-Rubio:2022ofy}. It is invariant under the subgroup of diffeomorphisms that preserve the four-dimensional volume element and contains the same number of propagating degrees of freedom as Einstein gravity. The cosmological constant $\Lambda_0$ does not appear in the unimodular action; instead, it enters the field equations as a free integration constant
~\cite{Unruh:1989,Henneaux:1989,VanDerBij:1982}. This decouples $\Lambda_0$ from the quantum vacuum energy of matter fields and reframes its value as an initial condition of the Universe rather than as a UV parameter~\cite{Anderson:1971,Buchmuller:1988a,Unruh:1989,Henneaux:1989,Weinberg:2000,Nobbenhuis:2006,Weinberg:1989}. When UG is combined with SI, $\Lambda_0$ enters the Einstein-frame scalar potential through the dynamically generated Planck scale and gives rise to a runaway exponential for the Goldstone boson of spontaneously broken scale symmetry, rather than to a constant vacuum-energy contribution~\cite{Cecchini:2026rou}. This mechanism was first demonstrated in~\cite{ Shaposhnikov:2008xb}, where a scale-invariant theory supplemented by the unimodular constraint was shown to yield a consistent cosmological history. It was subsequently developed into a complete Higgs--dilaton cosmology in~\cite{Garcia-Bellido:2011kqb,Garcia-Bellido:2012npk,Bezrukov:2012hx,Rubio:2014wta,Trashorras:2016azl,Casas:2017wjh,Rubio:2020zht,Piani:2022gon} and later analysed from the perspective of transverse diffeomorphism theories in \cite{Karananas:2016kyt,Casas:2018fum}.

In this paper, we introduce and analyse the unimodular extension of scale-invariant quadratic gravity~\cite{Rinaldi:2015,Ferreira:2019,Rinaldi:2023mdf,Cecchini:2024}. The central observation is that the Goldstone boson of spontaneously broken SI plays two sharply distinct cosmological roles. During inflation, conservation of the scale current freezes the Goldstone direction and leaves effectively single-field scalaron dynamics. At late times, the unimodular integration constant lifts this otherwise flat direction and turns the same field into dynamical DE. After canonical normalisation, the resulting exponential potential has a slope determined entirely by the field-space metric evaluated at the post-inflationary attractor. The late-time sector therefore contains no independent slope parameter: its dynamics are inherited directly from the inflationary geometry. We show that this geometry excludes the matter-era tracker branch throughout the parameter space and, within the inflationary viability window, automatically places the Goldstone field in the accelerating thawing regime~\cite{Caldwell:2005,Copeland:2006}. This behaviour is not restricted to the benchmark parameters considered below, but follows throughout the same parameter region that
supports successful inflation.

\begin{figure}
    \centering
    \includegraphics[width=0.7\linewidth]{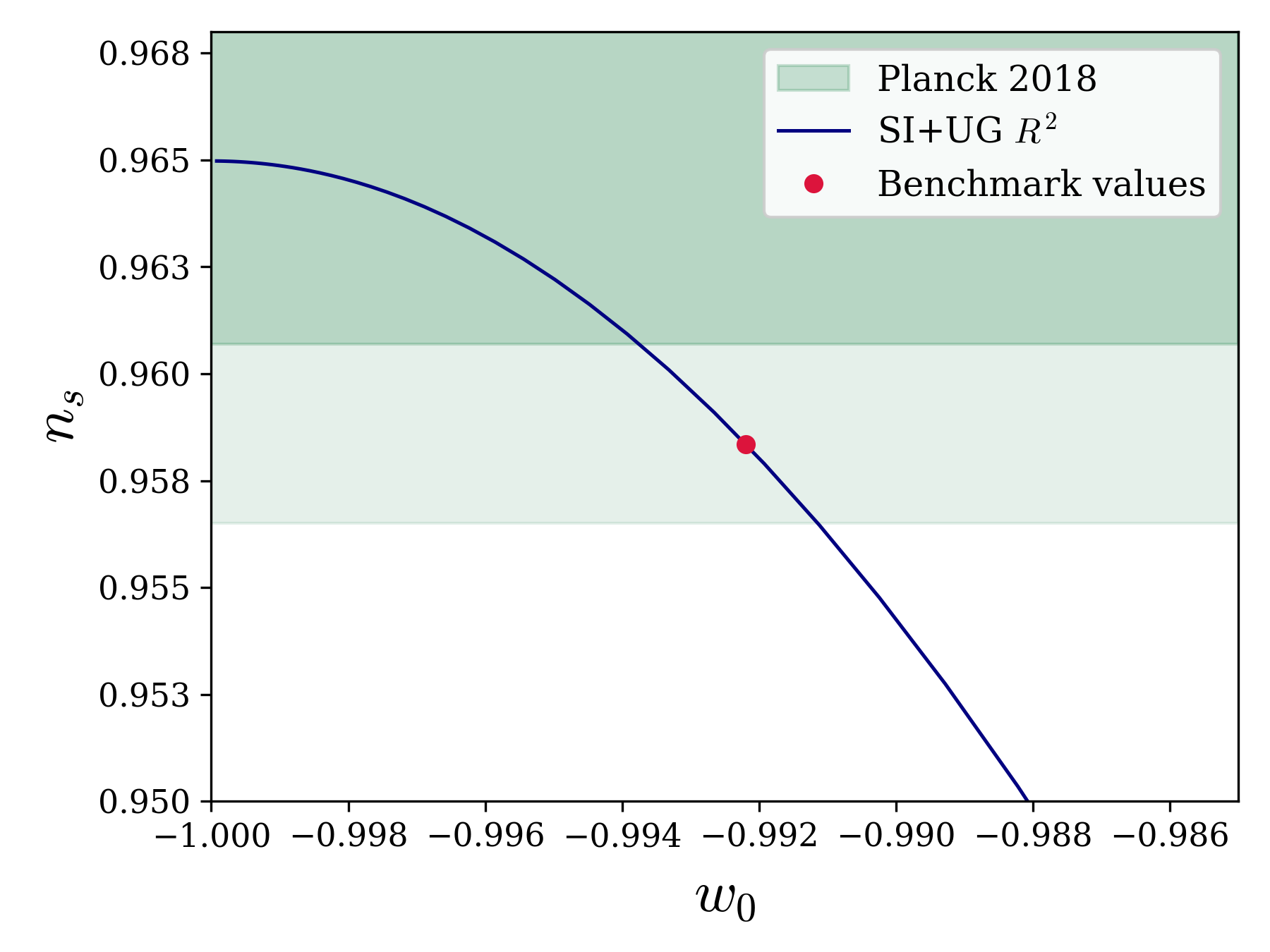}
    \caption{Consistency relation between the spectral index $n_s$ and the present DE equation-of-state parameter $w_0$ in the SI+UG $R^2$ model, obtained from \eqref{eq:intro_early_late_relation}. The green band shows the Planck 2018 $1\sigma$ and $2\sigma$ constraints on $n_s$. The red dot marks our benchmark point with $r = 2.87 \times 10^{-3} $.}
    \label{fig:ns_w0_plane}
\end{figure}

Most importantly, the common geometric origin of the inflationary and DE sectors gives rise to a \textit{direct consistency relation linking early- and late-Universe observables}, namely the spectral tilt $n_s$, the tensor-to-scalar ratio $r$, and the present-day DE equation-of-state parameter $w_0$ (see Fig.\,\ref{fig:ns_w0_plane}). At leading order, we find
\begin{equation}
(1-n_s)^2-\frac{r}{3}\simeq\left(\frac{3(1+w_0)}{2F(\Omega_{\rm DE})}\right)^2\,,
\label{eq:intro_early_late_relation}
\end{equation}
where \(F(\Omega_{\rm DE})\simeq0.45\) is a thawing-response function depending only on the present DE density fraction $\Omega_{\rm DE}\simeq0.69$. The combination on the left-hand side of \eqref{eq:intro_early_late_relation} vanishes in the pure Starobinsky limit, where $r\simeq3(1-n_s)^2$, whereas the right-hand side measures the present departure from a cosmological constant. For our representative benchmark, this gives \((w_0,w_a)\simeq(-0.992,-0.011)\) in the CPL parametrisation $w(a)=w_0+w_a(1-a)$~\cite{Chevallier:2001,Linder:2003}. The resulting thawing signal lies close to $\Lambda$CDM, while the early-late consistency relation makes the framework increasingly testable as measurements of $n_s$, $r$, and the DE equation-of-state parameter $w_0$ improve.

The present work is distinct from two related lines of research. The original realisations of scale-invariant unimodular cosmology employ the Standard Model (SM) Higgs as the inflaton \cite{Rubio:2019}, with an extra dilaton field providing the late-time DE~\cite{Shaposhnikov:2008xb,Garcia-Bellido:2011kqb,Casas:2018fum,Rubio:2017gty}. In those models, a consistency relation links the present equation-of-state parameter to the inflationary spectrum~\cite{Garcia-Bellido:2011kqb}, yielding a near-$\Lambda$CDM thawing prediction qualitatively similar to the one obtained here. The present construction realises an analogous early--late connection within a different inflationary framework: inflation is driven by the scalaron of the scale-invariant $R^2$ sector~\cite{Ferreira:2018,Ferreira:2019,Rinaldi:2023mdf,Cecchini:2024}, placing the dynamics squarely in the Starobinsky class with $r\sim10^{-3}$, while DE is carried by the pure scale Goldstone, decoupled from the SM. The consistency relation~\eqref{eq:intro_early_late_relation} then measures the thawing signal directly through the departure from the pure Starobinsky prediction in the $(n_s,r)$ plane.

The present framework also differs in kind from recent analyses of the DESI DE preference that combines non-parametric reconstructions of $w(z)$ and of the effective field theory of DE coefficients with concrete alternative-gravity proposals fitted to the data, such as the non-minimally coupled scalar-field (``\textit{thawing gravity}'') framework of~\cite{Ye:2025}. In those approaches, the late-time sector is either reconstructed without a fixed functional form or contains parameters adjusted to match observations. Here, by contrast, the potential and its slope are outputs of the inflationary attractor. The model therefore predicts which equation of state is selected by the symmetry, rather than which one is preferred by the data.

The paper is organised as follows. Section~\ref{sec:2} presents the scale-invariant quadratic-gravity model and its unimodular extension, deriving the Goldstone potential on the post-inflationary attractor. Section~\ref{sec:3} covers the early cosmology, including the inflationary trajectory, the CMB observables, and reheating through the Boltzmann equations. Section~\ref{sec:4} follows the post-reheating thermal history through radiation and matter domination, verifying that the quintessence field remains dynamically inert throughout these epochs. Section~\ref{sec:DE} presents the late-time evolution. We integrate the Goldstone equation of motion to the present time, derive the CPL parameters $w_0$ and $w_a$, compare the predictions with DESI constraints~\cite{Popovic:2025}, and establish the trade-off between the inflationary and DE sectors. Finally, we conclude in Section~\ref{sec:6}.

\section{The model}\label{sec:2}

In Sec.~\ref{sec:SI} we review the scale-invariant quadratic gravity model of~\cite{Rinaldi:2015,Rinaldi:2023mdf,Cecchini:2024}: starting from the Jordan-frame action built from dimension-four operators, we pass to the Einstein frame, identify the two-field sigma-model structure, and show how the Noether current of SI constrains the trajectory to an elliptic orbit, reducing the dynamics to a single effective degree of freedom. The vacuum structure of the theory is considered in Sec.~\ref{sec:lambda_naturalness}, and the dynamical stability of the post-inflationary fixed point is estabilished in Appendix\,\ref{sec:fixedpoint}. In Sec.~\ref{sec_themodel} we introduce the unimodular sector (whose construction is reviewed in Appendix~\ref{app:ug}) and show that the unimodular integration constant $\Lambda_0$ lifts this flat direction, producing an exponential potential for the Goldstone boson $\chi$ that acts as the DE field at late times.

\subsection{Scale-invariant quadratic gravity}\label{sec:SI}

We consider the scale-invariant gravitational action first studied in Refs.~\cite{Rinaldi:2015,Rinaldi:2023mdf} and subsequently confronted with CMB
data in \cite{Cecchini:2024}.  In the Jordan frame, it reads
\begin{equation}
  S_{J} = \int d^{4}x\,\sqrt{-g}
  \left[ \frac{\alpha}{36}\,R^{2} + \frac{\xi}{6}\,\phi^{2}R
    - \frac{1}{2}(\partial\phi)^{2}- \frac{\lambda}{4}\,\phi^{4}
  \right]\,,
\label{eq:SJ}
\end{equation}
where $\alpha$, $\xi$, and $\lambda$ are dimensionless coupling constants and $\phi$ is a real scalar field.  The action~\eqref{eq:SJ} is invariant under the global Weyl
rescaling $g_{\mu\nu}\to e^{2\sigma}g_{\mu\nu},\; \phi \to e^{-\sigma}\phi$,
for constant $\sigma$~\cite{Ferreira:2018,Ferreira:2019,Karananas:2021gco}, and invariant under coordinate dilatations $x\rightarrow e^{-\omega}x$, with $\omega=\rm const$ (for more details see \cite{DeAngelis:2025epa}). Every operator has mass dimension four and no explicit mass scale is present; all dimensionful quantities must therefore emerge dynamically through the spontaneous symmetry breaking of SI~\cite{Bardeen:1995kv,Meissner:2007,Wetterich:1988,Salvio:2014}. 

\begin{figure}
    \centering
    \includegraphics[width=\linewidth]{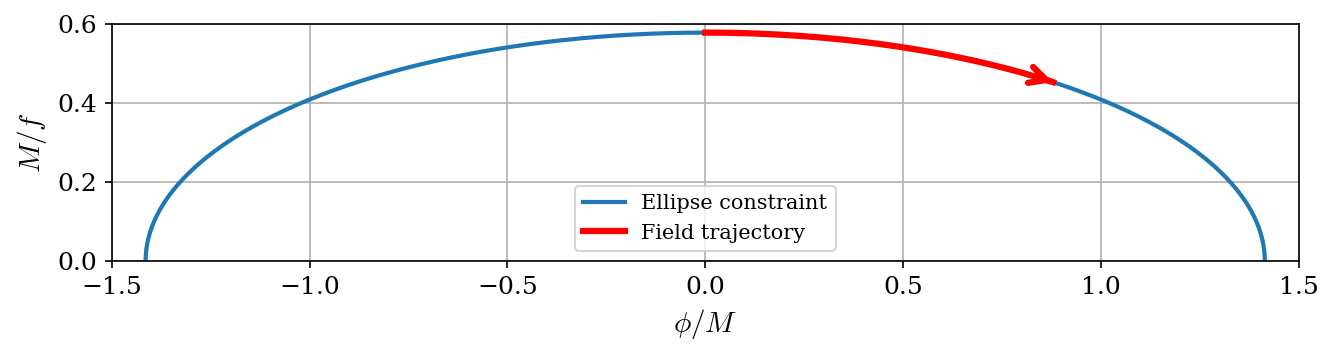}
    \caption{Elliptic orbit traced by the classical trajectory in the $(\phi,1/f)$ plane, set by conservation of the scale-symmetry Noether charge. Starting at $\phi=0$, the trajectory reduces the two-field dynamics to a single effective degree of freedom.}
    \label{fig:ellipse}
\end{figure}

The $R^{2}$ term is the unique ghost-free quadratic-curvature operator beyond Einstein--Hilbert in four dimensions and introduces a single propagating scalar (the scalaron)~\cite{Starobinsky:1980}; its coefficient $\alpha$ is fixed by the observed amplitude of the scalar power spectrum. Since the $R^2$ term contains fourth-order derivatives, it is convenient to reduce the action to second order by introducing an auxiliary field $\mathcal{F}$ through the identification
\begin{equation}
  \frac{\mathcal{F}^{2}}{M^2} = \frac{\alpha}{9M^2}\,R + \frac{\xi}{3M^2}\,\phi^{2}\,,
\label{eq:f}
\end{equation}
so that \eqref{eq:SJ} can be written as
\begin{equation}\label{Jframeuxiliar}
    S_{J}=\int d^4x\, \sqrt{-g}\left[\frac{\mathcal{F}^2}{2}R-\frac{1}{2}(\partial\phi)^2-\frac{9}{4\alpha}\mathcal{F}^4+\frac{3\xi}{2\alpha}\mathcal{F}^2\phi^2-\frac{\Omega}{4\alpha}\phi^4\right]\,, 
\end{equation}
with $\Omega\equiv \alpha\lambda+\xi^2$. In what follows it is convenient to work with the inverse of the auxiliary field, $f\equiv M^2/\mathcal{F}$, in terms of which the Einstein-frame quantities below take a simple form. The reference scale $M$ introduced above is arbitrary. SI implies that $M$ is dynamically identified with the integration constant of the conserved Noether current, as we now show. Indeed, SI gives rise to a conserved Noether current $K_{\mu} \equiv \partial_{\mu}K$,  with kernel~\cite{Garcia-Bellido:2011kqb,Ferreira:2016wem,Ferreira:2018,Ferreira:2018itt,Ferreira:2019,Cecchini:2024}
\begin{equation}
  K \;\equiv\; \frac{M^{2}}{2}
  \left(\frac{\phi^{2}}{M^{2}} + \frac{6M^{2}}{f^{2}}\right)\,,
\label{eq:K}
\end{equation}
which satisfies $\nabla_{\mu}K^{\mu} = 0$ along the equations of motion. The general solution to this conservation equation is \cite{Garcia-Bellido:2011kqb,Ferreira:2019}
\begin{equation}
  K = c_{1} + c_{2}\int\frac{dt}{a^{3}(t)}\,,
\label{eq:Ksol}
\end{equation}
so that $K$ rapidly approaches a constant value $c_{1}\neq 0$ in an expanding Universe, spontaneously breaking scale symmetry \cite{Garcia-Bellido:2011kqb,Ferreira:2016wem,Ferreira:2018itt}. Without loss of generality, we can set $c_{1} = M^{2}$, thereby establishing the structural mass scale $M$. Since the Weyl rescaling below promotes this constant to the Einstein-frame gravitational coupling, we will identify it with the reduced Planck mass $M\equiv M_P$. The conservation of $K$ confines the classical trajectory in the $(\phi,\,1/f)$ plane, which is constrained to lie on the ellipse ~\cite{Garcia-Bellido:2011kqb,Karananas:2016kyt,Cecchini:2024,Ferreira:2018,Ferreira:2019}
\begin{equation}
  f = \frac{\sqrt{6}\,M^{2}}{\sqrt{2M^{2} - \phi^{2}}}\,.
\label{eq:ellipse}
\end{equation}
This constraint reduces the two-field dynamics to a single effective degree of freedom (see Fig.~\ref{fig:ellipse}) and is directly responsible for the decoupling of entropy perturbations~\cite{Garcia-Bellido:2011kqb,Karananas:2016kyt,Cecchini:2024,Ferreira:2018,Ferreira:2019}. At the onset of inflation ($\phi = 0$), the kernel $K$ simplifies to $3M^{4}/f^{2}$, yielding the initial high-energy value $f_{\text{saddle}}=\sqrt{3}M$. Furthermore, the geometric constraint imposed by the conserved scale-symmetry Noether current forces $f$ to increase (equivalently, the auxiliary field $\mathcal{F}$ to decrease) monotonically as $\phi$ evolves from the inflationary saddle $\phi=0$ towards the stable minimum of the potential. 
As a result, $f$, and hence the auxiliary field $\mathcal{F}$, acquires a dynamically generated vacuum expectation value (VEV). Remarkably, this scale is entirely determined by the dimensionless inflationary couplings,
\begin{equation}
    \frac{\langle f\rangle}{M_P} =\sqrt{3\left(1+\frac{\xi}{2\Omega}\right)}\,.
    \label{eq:VEV}
\end{equation}
Although $\langle f\rangle$ is numerically larger than $M_P$, the Einstein-frame gravitational coupling remains fixed by $M_P$. The Jordan-frame late-time Planck scale, dynamically selected by the inflationary attractor, is instead set by the auxiliary field, $\langle \mathcal{F}\rangle=M_P^2/\langle f\rangle$, which for our benchmark parameters gives $\langle \mathcal{F}\rangle\simeq0.08\,M_P$. 

A Weyl transformation $\tilde{g}_{\mu\nu} =(M_P^{2}/f^2)\,g_{\mu\nu}$ then brings the model \eqref{Jframeuxiliar} into the Einstein frame~\cite{Rinaldi:2015,Rinaldi:2023mdf,
Cecchini:2024}. For $f\neq 0$,  the action can then be written in the compact form of a two-field non-linear
sigma model~\cite{Cecchini:2024,Ferreira:2018,
Ferreira:2019},
\begin{equation}
  S_{E} = \int d^{4}x\,\sqrt{-\tilde{g}}
  \left[   \frac{M_P^{2}}{2}\tilde{R}- \frac{1}{2}\,\mathcal{G}_{IJ}\,\tilde{g}^{\mu\nu}
      \partial_{\mu}\phi^{I}\partial_{\nu}\phi^{J}- V(\phi^{I}) \right],
  \quad \phi^{I} \equiv \binom{\phi}{f}\,,
\label{eq:SE_phif}
\end{equation}
with field-space metric~\cite{Cecchini:2024}
\begin{equation}
  \mathcal{G}_{IJ} =
  \begin{pmatrix}
    e^{2b(f)} & 0 \\[4pt]
    0 & 6\,e^{-2b(f)}
  \end{pmatrix}\,,
  \qquad
  b(f) \equiv \ln\!\left(\frac{f}{M_P}\right),\,
\label{eq:GIJ}
\end{equation}
and scalar potential~\cite{Cecchini:2024}
\begin{equation}
  V(\phi, f) = -\frac{3\xi}{2\alpha}\,\phi^{2}f^{2} + \frac{\Omega}{4\alpha M_P^{4}}\,\phi^{4}f^{4} + \frac{9M_P^{4}}{4\alpha}\,.
\label{eq:Vphif}
\end{equation}
It is convenient to perform the field redefinition introduced in Ref.~\cite{Cecchini:2024},
\begin{equation}
  \rho = \sqrt{6}\,M_P\,\mathrm{arcsinh} \!\left(\frac{\phi f}{\sqrt{6}\,M_P^{2}}\right)\,,
  \qquad
  \chi = \frac{M_P}{2}\ln\!\left(\frac{\phi^{2}}{2M_P^{2}} + \frac{3M_P^{2}}{f^{2}}
  \right)\,.
\label{eq:fields}
\end{equation}
In terms of these variables, the action~\eqref{eq:SE_phif} takes the form
\begin{equation}
  S_{E} = \int d^{4}x\,\sqrt{-\tilde{g}}\left[ \frac{M_P^{2}}{2}\tilde{R}
    - \frac{1}{2}(\partial\rho)^{2} - \frac{1}{2}\,e^{2b(\rho)}(\partial\chi)^{2}- V(\rho) \right]\,,
\label{eq:SE}
\end{equation}
where $\rho$ carries a canonical kinetic term and the field-space metric
coefficient for $\chi$ is given by
\begin{equation}
  b(\rho) = \ln\!\left(\sqrt{6}\cosh\frac{\rho}{\sqrt{6}\,M_P}\right).
\label{eq:b}
\end{equation}
The Einstein-frame potential now reads~\cite{Rinaldi:2015,
Rinaldi:2023mdf,Cecchini:2024},
\begin{equation}
  V(\rho) = \frac{9M_P^{4}}{4\alpha}
  \left[1 - 4\xi\sinh^{2}\!\left(\frac{\rho}{\sqrt{6}\,M_P}\right) + 4\Omega\sinh^{4}\!\left(\frac{\rho}{\sqrt{6}\,M_P}\right) \right].
\label{eq:Vrho}
\end{equation}
This potential belongs to the same universality class as pole-inflation and \(\alpha\)-attractor models~\cite{Galante:2014ifa,Artymowski:2016pjz,Karananas:2016kyt}. Note that prior to the inclusion of the unimodular contribution, it is completely independent of $\chi$. This field is the Goldstone boson associated with the spontaneous breaking of the global scale symmetry and is therefore exactly massless at tree level~\cite{Shaposhnikov:2008xb,Garcia-Bellido:2011kqb,Ferreira:2018,Karananas:2016kyt,Karananas:2016grc}. Despite being massless, it need not mediate an observable long-range fifth force~\cite{Shaposhnikov:2008xb,Garcia-Bellido:2011kqb,Ferreira:2016kxi,Karananas:2016grc}.~\footnote{Or, in Han Solo's words: ``\textit{That's not how the Force
works!''}} As we shall see below, the unimodular integration constant lifts the otherwise flat direction, generating the exponential potential responsible for the late-time DE dynamics. Before introducing this lifting, however, it is instructive to examine the vacuum structure of the scale-invariant sector itself. 

\subsection{Inflationary attractor and Minkowski vacuum}
\label{sec:lambda_naturalness}

Since the scale-invariant potential~\eqref{eq:Vrho} is independent of the Goldstone field $\chi$, its non-trivial stationary condition defines a one-dimensional vacuum manifold,
\begin{equation}
\mathcal M_{\rm vac}=\left\{(\rho,\chi)\,:\,\rho=\rho_{\min}\,,\ \chi\in\mathbb R\right\}\,,\qquad\textrm{with}\qquad  \sinh^2 \frac{\rho_{\rm min}}{\sqrt{6}M_P}=\frac{\xi}{2\Omega}\,. 
\label{eq:vacuum_manifold}
\end{equation}
The continuous degeneracy along $\chi$ follows from the spontaneous breaking of SI and persists independently of the value of $\lambda$. This coupling, therefore, does not determine the existence of the Goldstone direction, but rather the vacuum energy associated with the entire manifold. Evaluating the potential at the radial minimum $\rho_{\rm min}$, we obtain
\begin{equation}
V(\rho_{\min})=\frac{9M_P^4}{4\alpha}\left(1-\frac{\xi^2}{\Omega}\right)=\frac{9M_P^4}{4}\frac{\lambda}{\Omega}\,.
\label{eq:vacuum_energy_lambda}
\end{equation}
Accordingly, $\lambda>0$, $\lambda=0$, and $\lambda<0$ correspond, respectively, to de Sitter, Minkowski, and anti-de Sitter branches of the scale-invariant vacuum manifold. The limit $\lambda\to 0$ ($\Omega\to\xi^2$) is nevertheless dynamically regular, as shown explicitly in Appendix~\ref{app:lambda0}.

The Minkowski branch, $\lambda=0$, is distinguished for several reasons. First, it is the unique branch for which the scale-invariant sector leaves no residual vacuum energy after inflation. A further consideration concerns the quantum behaviour of the Goldstone mode. Spontaneous breaking of SI implies the existence of an exactly massless scalar degree of freedom, irrespective of the value of $\lambda$. For $\lambda\neq0$, however, this Goldstone boson propagates on a de Sitter or anti-de Sitter background. Massless scalar fields are known to exhibit infrared instabilities in de Sitter spacetime \cite{Allen:1987tz}, while related subtleties have also been suggested in four-dimensional anti-de Sitter space \cite{Bizon:2011gg}. It is therefore conceivable that a consistent quantisation of the spontaneously broken scale-invariant theory may favour the Minkowski branch. This provides an additional theoretical motivation for the condition $\lambda=0$, along the lines previously advocated in the Higgs--Dilaton framework \cite{Shaposhnikov:2008xb,Garcia-Bellido:2011kqb} (see also \cite{Antoniadis:1985pj,Tsamis:1992sx,
Tsamis:1994ca,Antoniadis:2006wq,Polyakov:2009nq}). 

These structural motivations should not, however, be confused with technical naturalness. The operator $\lambda\phi^4$ is fully compatible with SI, and setting its coefficient to zero does not restore any additional manifest symmetry of the action. Radiative corrections may therefore generate a non-vanishing effective quartic coupling and shift the theory away from $\Omega=\xi^2$. At the quantum level, the Minkowski condition should consequently be understood as a renormalisation condition imposed on the effective potential, $V_{\rm eff}(\rho_{\min})=0$. The Minkowski branch, therefore, remains a tuned choice at the quantum level. More generally, implementing scale or conformal symmetry at the quantum level requires a symmetry-compatible regularisation prescription, and its relation to the gravitational Weyl anomaly is subtle \cite{Shaposhnikov:2008xi,Shaposhnikov:2008ar,Monin:2013gea,Armillis:2013wya,Bezrukov:2014ipa,Shaposhnikov:2018nnm,Mooij:2018hew,Shaposhnikov:2022dou,Shaposhnikov:2022zhj}.

On the de Sitter branch, the residual vacuum energy in Eq.~\eqref{eq:vacuum_energy_lambda} could itself account for the observed dark-energy density without invoking a unimodular sector, as discussed in Ref.~\cite{Cecchini:2026rou}. This possibility yields a cosmological-constant contribution with $w=-1$, but does not generate the dynamical early--late connection
studied here. Also, neither this alternative nor our construction explains the smallness
of the observed dark-energy scale. In the former, the residual vacuum
energy is adjusted through the dimensionless coupling $\lambda$.
In the latter, the Minkowski condition is imposed on the
scale-invariant effective potential, while the unimodular integration
constant fixes the amplitude of the late-time potential.
Our reason for selecting the Minkowski branch is therefore its
predictive structure: the dark-energy component originates entirely
from the unimodular lifting, and its slope is determined by the same
parameters that govern inflation.
We restrict the remainder of the analysis to $\lambda=0$, or
equivalently $\Omega=\xi^2$, except when explicitly discussing the
effect of a nonzero residual vacuum energy.

\subsection{Unimodular lifting of the Goldstone direction}
\label{sec_themodel}

Having selected the Minkowski branch of the scale-invariant vacuum manifold, we now introduce the ingredient that lifts the otherwise flat Goldstone direction. In UG, the metric determinant is constrained to a fixed value, while the cosmological constant is absent as a parameter of the action and instead emerges as an integration constant $\Lambda_0$ of the gravitational field equations. When combined with SI, this integration constant does not generate an ordinary constant vacuum-energy contribution in the Einstein frame. Rather, it couples to the dynamically generated conformal factor and induces a runaway potential along the Goldstone direction. The resulting lifting of the Minkowski vacuum manifold provides the late-time DE sector of the model. The unimodular construction is reviewed in Appendix~\ref{app:ug}.

The combined theory is obtained by supplementing the scale-invariant Jordan-frame action~\eqref{eq:SJ} with the unimodular constraint, implemented through a Lagrange multiplier $\Lambda(x)$,
\begin{equation}
  S = \int d^{4}x\left\{  \sqrt{-g}\left[\frac{\alpha}{36}\,R^{2} + \frac{\xi}{6}\,\phi^{2}R  - \frac{1}{2}(\partial\phi)^{2} - \frac{\lambda}{4}\,\phi^{4}
    \right] + \Lambda(x)\!\left[\sqrt{-g} - 1\right]
  \right\}\,.
\label{eq:Sfull}
\end{equation}
Although we retain $\lambda$ in Eq.~\eqref{eq:Sfull} to display the general
structure of the theory, the following analysis is specialised
to the Minkowski branch $\lambda=0$, except for the discussion
of a nonzero residual vacuum energy in Sec.~\ref{sec:early-late}.

Following~\cite{ Shaposhnikov:2008xb,Garcia-Bellido:2011kqb,Casas:2018fum}, the unimodular integration constant enters the Einstein-frame scalar potential through the conformal factor $f$ according to
\begin{equation}
  V_{\rm UG}(f) = \Lambda_{0}\,\frac{f^{4}}{M_P^{4}}\,.
\label{eq:VUG}
\end{equation}
The total Einstein-frame scalar potential is therefore $
V_{\rm tot}(\phi, f) = V(\phi,f) +   V_{\rm UG}(f)$,  where $V(\phi,f)$ is given by \eqref{eq:Vphif}. The quartic dependence of \(V_{\rm UG}\) on \(f\) is fixed by the combined SI+UG structure \cite{ Shaposhnikov:2008xb,Garcia-Bellido:2011kqb,Casas:2018fum}.  Once restricted to the post-inflationary vacuum manifold, this dependence lifts the otherwise flat Goldstone direction and generates the non-trivial runaway potential that drives the late-time DE dynamics.
 
It is worth stressing the distinct roles played by each ingredient of the combined theory. The contracted Bianchi identity $\nabla^{\mu}G_{\mu\nu} = 0$ holds as a differential identity for any metric and any gravitational Lagrangian, including the $R^{2}$ theory~\cite{Padilla:2015,Jirousek:2023,Carballo-Rubio:2022ofy}.
The emergence of $\Lambda_0$ as a free integration constant through \eqref{eq: lambda_const} is therefore a feature of the unimodular constraint alone, and is not altered by the presence of the $R^{2}$ term in the action. The two frameworks are complementary: via the auxiliary field $\mathcal{F}$ and the Weyl transformation, the scalaron acts as the inflaton and drives the early-Universe dynamics, while the unimodular term provides the potential tail for the Goldstone boson $\chi$ that sources the late-time DE. To obtain the late-time potential for the Goldstone mode, we evaluate the effective potential on the inflationary attractor, where $\rho$ has settled at the post-inflationary minimum $\rho_{\min}$ (see Appendix\,\ref{sec:fixedpoint}). The field redefinition \eqref{eq:fields} yields two simultaneous relations among $(\phi,1/f)$,
\begin{equation}
  \phi\,f = \sqrt{6}\,M_P^{2}\,\sinh(u_{\min})\,,
  \qquad
  e^{2\chi/M_P}=\frac{\phi^{2}}{2M_P^{2}} \,+\, \frac{3M_P^{2}}{f^{2}} \,,
  \label{eq:conditions}
\end{equation}
with $u\equiv \rho/(\sqrt{6}M_P)$. Eliminating $\phi$ with the first equation in \eqref{eq:conditions} yields
$e^{2\chi/M_P} = (3M_P^{2}/f^{2})\cosh^{2}(u_{\min})$, hence the exact relation
\begin{equation}
  f^{2} = 3\,M_P^{2}\,\cosh^{2}(u_{\min})\,e^{-2\chi/M_P} \,,
  \label{eq:f-of-chi}
\end{equation}
which holds for arbitrary $\chi$. Substituting into~\eqref{eq:VUG} gives the exact unimodular potential along the attractor,
\begin{equation}
  V_{\rm UG}(\chi) = \frac{\Lambda_{0}\,\langle f\rangle^{4}}{M_P^{4}}\,e^{-4\chi/M_P} \,,
\end{equation}
where $\langle f\rangle = \sqrt{3}\,M_P\cosh\,u_{\min}$ is the VEV defined in~\eqref{eq:VEV} and the overall normalisation is fixed by matching it to the present DE density, 
\begin{equation}\label{Lambda_obs}
V_{\rm DE}^{(0)}=\frac{\Lambda_0\langle f\rangle^4}{M_P^4}
\simeq 10^{-120}M_P^4\,.
\end{equation}
Once $\rho$ has relaxed to $\rho_{\rm min}$, the coefficient $e^{b(\rho_{\rm min})}$ becomes a constant and the $\chi$ sector acquires an effective exponential DE potential through $V_{\rm UG}$ \cite{Rubio:2017gty}. Indeed, we can define the canonically normalised \textit{Goldstone} by
\begin{equation}
    \Psi\equiv e^{b(\rho_{\rm min})}\chi, \qquad \quad b(\rho_{\rm min}) = \dfrac{1}{2}\ln \left [ 6\cosh^2 \left ( u_{\rm min} \right ) \right ] \,,
    \label{eq: psgold}
\end{equation}
so that the unimodular potential \eqref{eq:VUG} takes the form
\begin{equation}
  V_{\rm UG}(\Psi) = V_{\rm DE}^{(0)}  e^{-\frac{4\gamma}{M_{P}}\,\Psi}\,.
\label{eq:Vexp}
\end{equation}
The late-time slope $\gamma$ is entirely determined by the inflationary sector parameters $\xi$ and $\Omega$ through the geometry of the fixed point,
\begin{equation}
\gamma \equiv e^{-b(\rho_{\rm min})}
= \sqrt{\frac{\Omega}{3(2\Omega+\xi)}}\,.
\label{eq:muPsi}
\end{equation}
Varying the pivot scale between $N_{*}=55$ and $N_{*}=60$ rescales $\alpha$ by
$\sim\!22\%$ to match the observed amplitude $A_{s}$, but leaves the field-space geometry at $\rho_{\rm min}$ (and hence $\gamma$) entirely unchanged at leading order.
Late-time DE constraints require $\gamma< 1/2\sqrt{2}$, which translates to
\begin{equation}
  \xi > \frac{2\Omega}{3}\,.
\label{eq:DEconstraint}
\end{equation}
For our parameter values $\xi=0.01$, $\Omega=10^{-4}$, which satisfy $\Omega=\xi^2$ (i.e.~$\lambda=0$, the choice that removes the residual vacuum energy), one finds $\gamma\simeq 0.058$, so that \eqref{eq:DEconstraint} is satisfied
and the field is deep in the thawing regime. Indeed, since $\cosh u_{\rm min}\geq 1$ for any $\xi,\Omega>0$, the model imposes the universal upper bound
\begin{equation}
  \gamma \;\leq\; \frac{1}{\sqrt{6}} <\; \frac{\sqrt{3}}{4}\,,
  \label{eq:mubound}
\end{equation}
with equality only in the unphysical limit $u_{\rm min}\to 0$. \footnote{This corresponds to the unphysical limit $\xi/\Omega \to 0$ (vanishing non-minimal coupling): the de Sitter saddle and the IR fixed point coincide at $\rho = 0$, so the inflaton potential admits no nontrivial minimum and the field never rolls.} Note that this bound holds for any $\xi, \Omega >0$, independently of $\lambda$; the benchmark above is simply the $\lambda=0$ point on this family.  For exponential potentials, a tracker attractor in the matter era requires $\gamma>\sqrt{3}/4$, while late-time accelerated expansion requires $\gamma<1/2\sqrt{2}$ \cite{Copeland:1997et,Ferreira:1998,Zlatev:1999,Steinhardt:1999,Caldwell:2005}. The bound \eqref{eq:mubound} therefore excludes the tracker branch structurally, for all admissible $(\xi,\Omega)$. This bound by itself, however, only enforces $\gamma<\sqrt{3}/4$; the stronger requirement $\gamma<1/2\sqrt{2}$ for accelerated expansion is equivalent to the condition \eqref{eq:DEconstraint}, which is not automatic for arbitrary $(\xi,\Omega)$. It is, however, guaranteed throughout the inflationary viability window, $\xi^{2}\leq\Omega\leq\tfrac{2\sqrt{3}}{3}\xi^{2}$ (see Appendix~\ref{app:lambda0}, where we show that the boundary $\Omega=\xi^{2}$, excluded by assumption in~\cite{Ghoshal:2022qxk}, is dynamically regular). Here, $2\Omega/3\leq\tfrac{4}{3\sqrt{3}}\xi^{2}<\xi$ for all $\xi<3\sqrt{3}/4\approx1.30$, so the entire physically admissible region with $\xi\leq 1$ lies in the thawing window $\gamma <1/(2\sqrt{2})$~\cite{Caldwell:2005,Scherrer:2006,Dutta:2008qn,Linder:2008}, with our benchmark $\xi=10^{-2}$ deep inside it.

\section{Early Universe}\label{sec:3}

In this section, we follow the evolution from inflation to the onset of radiation domination. We first compute the inflationary observables along the Noether attractor and then model the perturbative transfer of the scalaron energy into a radiation bath, determining the reheating temperature \(T_{\rm reh}\) \cite{Barman:2025lvk}.

\subsection{Inflationary predictions}\label{subsec:infl}

Starting from the action~\eqref{eq:SE}, the equations of motion in a flat Friedmann--Lema\^{i}tre--Robertson--Walker background are \cite{Ferreira:1998,Ferreira:2019,Cecchini:2024}
\begin{equation}
  \ddot{\rho} + 3H\dot{\rho} + V_{,\rho}- b_{,\rho}\,e^{2b(\rho)}\dot{\chi}^{2} = 0\,, \qquad \quad 
  \ddot{\chi} + 3H\dot{\chi} + 2b_{,\rho}\,\dot{\rho}\dot{\chi} = 0\,,
    \label{eq:EOM_rho}
\end{equation}
closed by the Friedmann equation
\begin{equation}
  H^{2} =\frac{1}{3 M_P^2} \bigg(\frac{\dot{\rho}^{2}}{2}
    + e^{2b(\rho)}\frac{\dot{\chi}^{2}}{2} + V(\rho)\bigg)\,.
  \label{eq:friedmann}
\end{equation}
Note that during inflation the unimodular contribution \eqref{eq:VUG} is negligible.  
\begin{figure}
    \centering
    \includegraphics[width=0.47\linewidth]{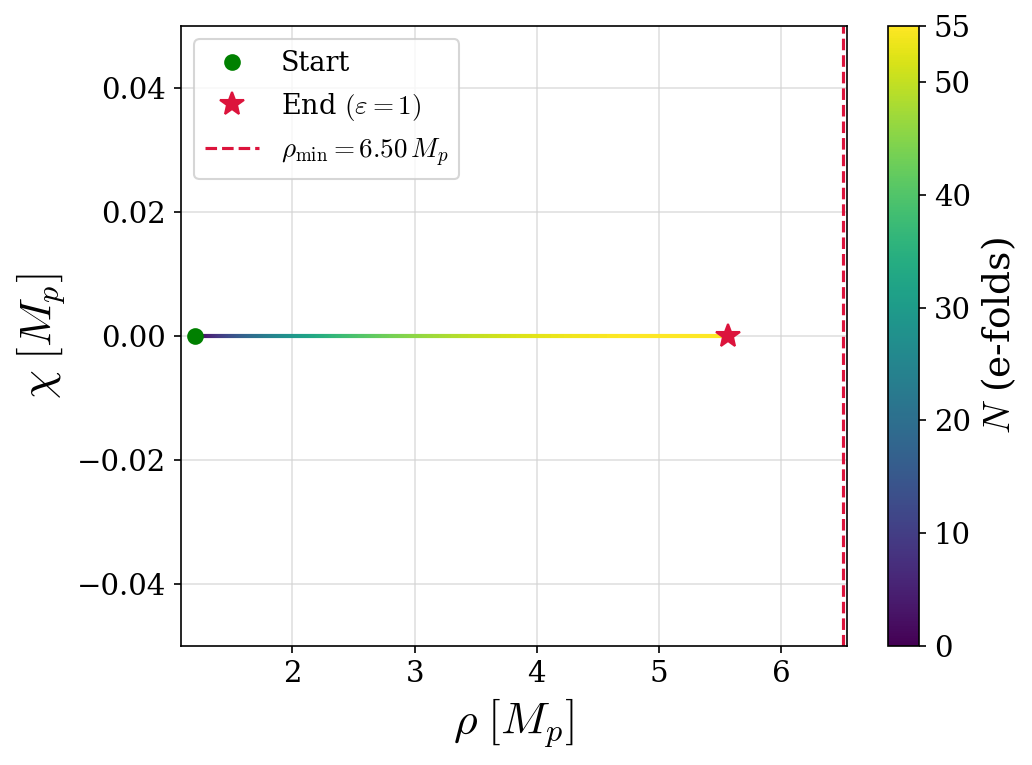}
    \hfill 
    \includegraphics[width=0.50\linewidth]{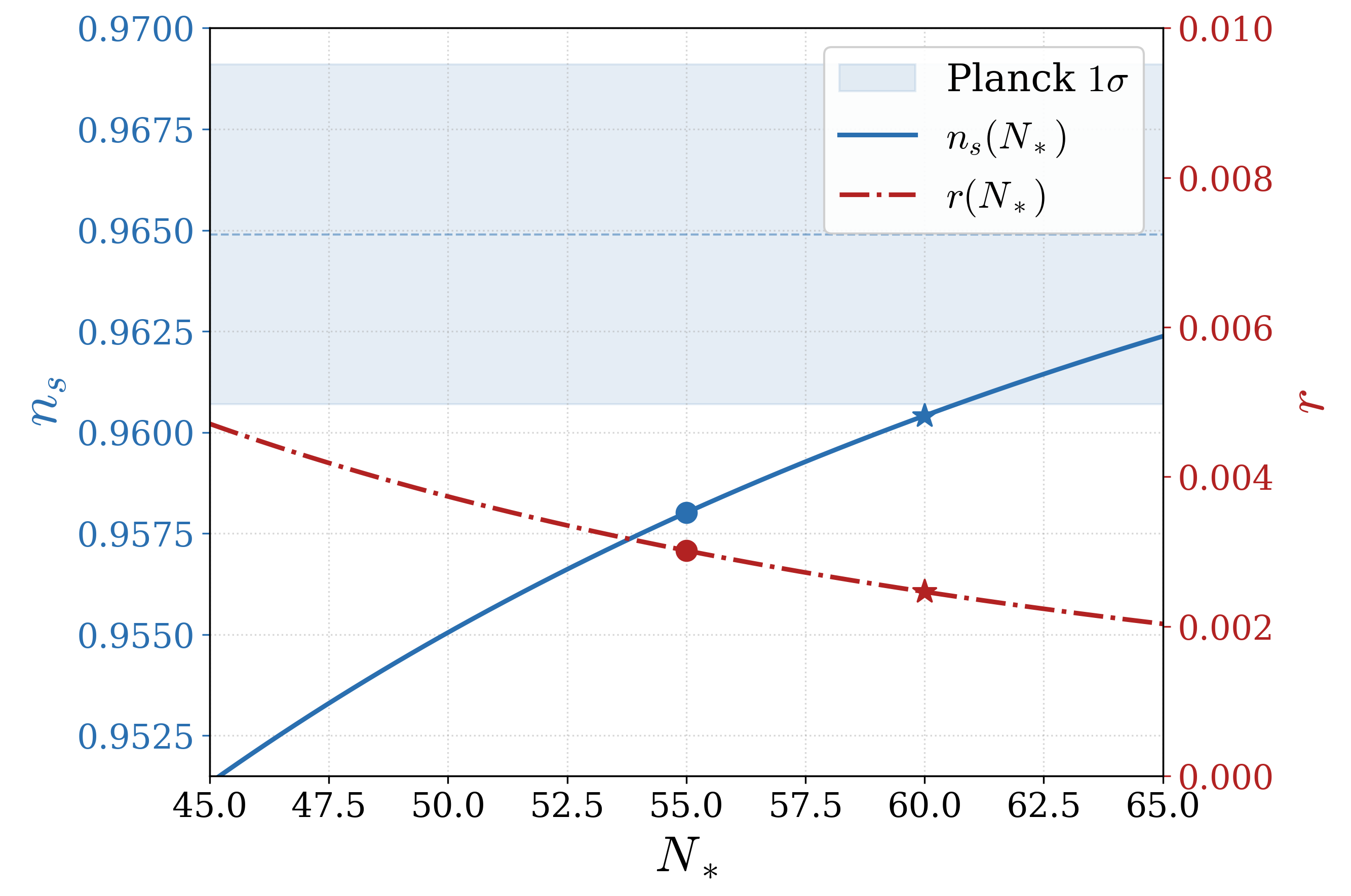}
    \caption{\textbf{Left}: Inflationary trajectory for $\alpha=2.496 \times 10^{10}$, $\xi=0.01$, $\Omega= 10^{-4}$. $\rho_{\rm min}$ marks the IR-stable fixed point (Appendix\,\ref{sec:fixedpoint}). \textbf{Right}: Spectral index and tensor-to-scalar ratio compared to Planck and BK18 bounds. We show $N_*=55$ and $N_*=60$, where for the latter $\alpha$ is increased by $\sim 22\%$ ($3.056\times10^{10}$) to keep $A_s$ within observed limits.}
    \label{fig:inflation_results}
\end{figure}
The shift symmetry $\chi\to\chi+c$ of the action~\eqref{eq:SE} (under which neither $V(\rho)$ nor $b(\rho)$ changes) is the Einstein-frame realisation of SI, and its conservation law is the second expression in \eqref{eq:EOM_rho}. The associated conserved current is precisely the current $K^\mu=\partial^\mu K$ introduced in Sec.~\ref{sec:SI}. As shown in \eqref{eq:Ksol}, $K$ rapidly approaches a constant in an expanding Universe, so $\chi$ freezes on the orbit $\chi=0$ irrespective of its initial value, cf.~\eqref{eq:ellipse}. With $\chi$ frozen at zero, the entropic mode does not source curvature perturbations along the Noether attractor,~\footnote{Even though $\chi$ is a massless spectator and acquires fluctuations $\delta\chi\sim H_{\rm infl}/2\pi$, these can at most survive as uncorrelated isocurvature and remain harmless here, since $\delta V/V\sim4\gamma\,\delta\Psi/M_P\sim10^{-7}$.} and the two-field system reduces to effective single-field dynamics~\cite{Garcia-Bellido:2011kqb,Ferreira:2018,Ferreira:2019} (see Fig.~\ref{fig:inflation_results}). The field $\rho$ rolls from the unstable de Sitter saddle at $\rho=0$ toward the stable minimum, $\sinh^{2}(\rho_{\rm min}/\sqrt{6}M_P) = \xi/(2\Omega)$ ~\cite{Rinaldi:2023mdf,Cecchini:2024} (see Appendix\,\ref{sec:fixedpoint}).  
The CMB observables are then computed in the slow-roll approximation for the single field $\rho$. The potential slow-roll parameters are
\begin{equation}
  \epsilon_{V} = \frac{M_P^{2}}{2}\left(\frac{V'}{V}\right)^{2}\,,
  \qquad
  \eta_{V} = M_P^{2}\frac{V''}{V}\,,
\label{eq:srpars}
\end{equation}
where primes denote derivatives with respect to $\rho$. The spectral index
and tensor-to-scalar ratio, 
\begin{equation}
  n_{s} = 1 - 6\epsilon_{V} + 2\eta_{V}\,,
  \qquad
  r = 16\epsilon_{V}\,,
\label{eq:nsrSR}
\end{equation}
are evaluated at the horizon-crossing field value $\rho_{i}(\xi,\Omega,N_{*})$,
defined implicitly by
\begin{equation}
  N_{*} = \frac{1}{M_P^{2}}\int_{\rho_{i}}^{\rho_{\rm end}} \frac{V}{V'}\,d\rho\,,
\label{eq:Nstar}
\end{equation}
where subscript \(\ast\) here indicates horizon crossing and $\rho_{\rm end}$ is determined by $\epsilon_{V}(\rho_{\rm end})=1$ \cite{Liddle:1992,Lyth:1999}.  For
the benchmark values $\xi=0.01$, $\Omega=10^{-4}$, numerical evaluation of~\eqref{eq:nsrSR}--\eqref{eq:Nstar} gives
\begin{equation}
  n_{s} = 0.9584\,,\quad r = 2.87 \times 10^{-3} \quad (N_{*}=55)\,,
  \qquad
  n_{s} = 0.9607\,,\quad r = 2.35 \times 10^{-3} \quad (N_{*}=60)\,.
\label{eq:CMBvals}
\end{equation}
The benchmark value $n_s=0.9584$ lies within $1.5\sigma$ of the Planck 2018 value, $n_s=0.9649\pm0.0042$~\cite{Planck:2018vyg,Planck:2020inf}, while the predicted tensor-to-scalar ratio,
$r\equiv r_{0.05}=2.87 \times 10^{-3}$, is well below current bounds on primordial
tensors~\cite{BICEP:2021,ACT:2025ext}. This comparison, however, refers to parameter constraints obtained under the cosmological models adopted in those analyses, and is therefore sensitive to the dataset and cosmological assumptions used in the inference.
Planck and ACT DR6 are mutually consistent when compared directly: ACT combined with WMAP gives $n_s=0.9660\pm0.0046$, against $0.9651\pm0.0044$ for Planck, agreeing to $0.2\sigma$~\cite{ACT:2025}. However, the preferred value of the scalar tilt shifts upward when ACT is combined with Planck, and increases further upon adding CMB lensing, DESI BAO, and BK18 data. The P-ACT combination gives $n_s=0.9709\pm0.0038$, while the public P-ACT-LB-BK18 chain~\cite{ACT:2025chains}, which includes DESI BAO, yields $n_s=0.9741\pm0.0033$. Relative to the latter combination, the benchmark is disfavoured at more than $4.8\sigma$. This upward shift has motivated extensions of Starobinsky inflation involving higher-curvature corrections or related deformations designed to raise the predicted scalar tilt \cite{Addazi:2025qra,Ketov:2025cqg,Gialamas:2025ofz,Qiu:2025iqm}.

The interpretation of the P-ACT-LB-BK18 combination nevertheless requires some caution. As emphasised in Ref.~\cite{Ferreira:2025lrd}, BAO data do not constrain $n_s$ directly: the upward displacement results from correlations within the CMB likelihood combined with the unresolved tension between the CMB and DESI determinations of the matter density parameter $\Omega_m$ and the sound horizon $r_dh$ under $\Lambda$CDM. CMB-only determinations remain mutually consistent and substantially closer to the Planck value. We therefore use the joint Planck 2018 + BK18 contours as a transparent CMB reference for displaying the intrinsic inflationary predictions of the model, while keeping in mind the stronger, but more dataset-dependent, shifts obtained from combinations involving DESI.

\begin{figure}
    \centering
    \includegraphics[width=0.8\linewidth]{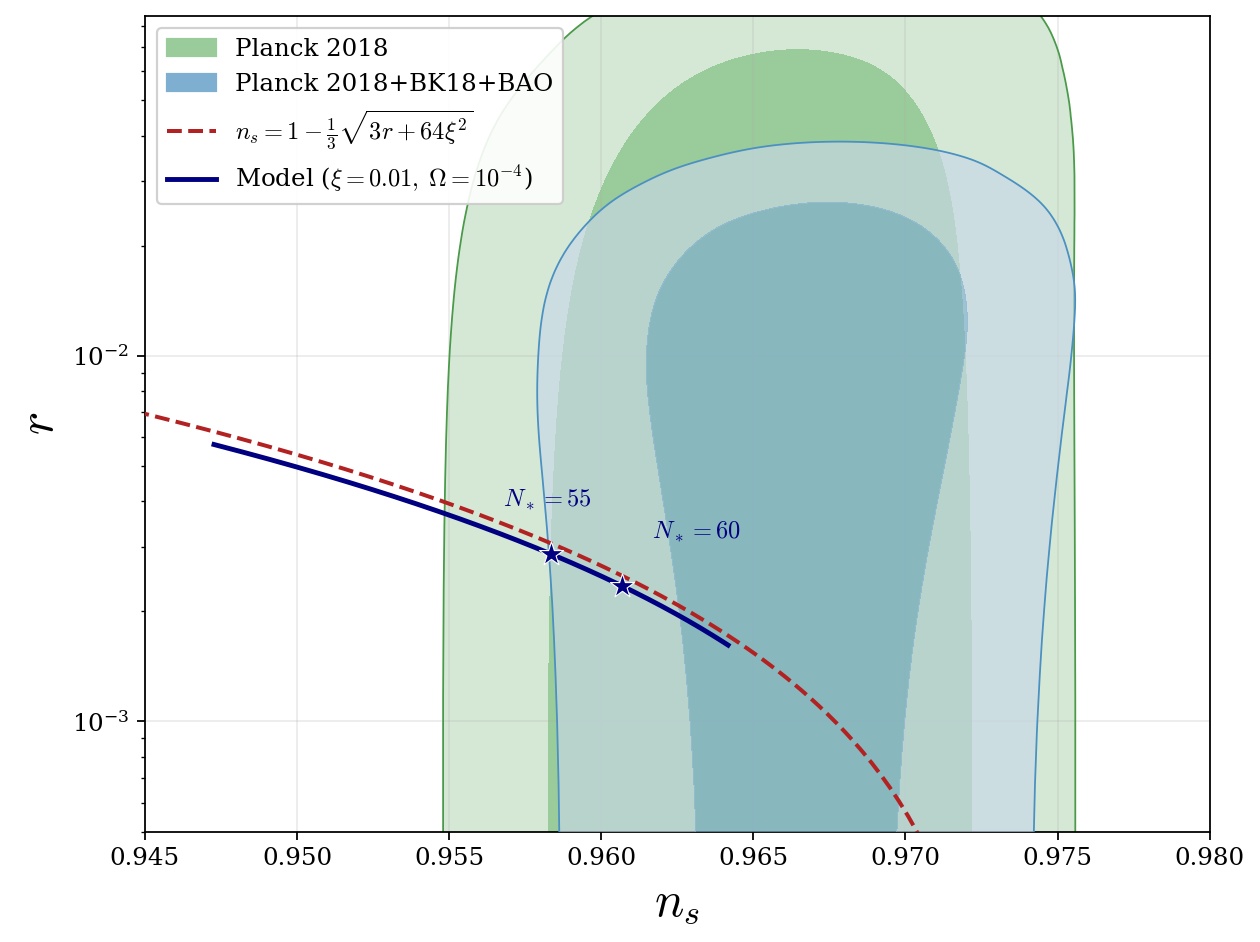}
    \caption{Constraints on the scalar spectral index $n_s$ and the tensor-to-scalar ratio $r$ from Planck\,2018 TT,TE,EE+lowE+lensing (green shaded contours) and from its combination with BK18 and BAO data (shaded blue contours) at 68\% and 95\% C.L. taken from the official BICEP/Keck data release~\cite{BICEP:2021}.  The solid navy curve is the model prediction as $N_*$ ranges from 40 to 70; stars mark $N_*=55$ and $N_*=60$. The dashed red curve is the analytic approximation \eqref{eq:nsranal}.}
    \label{fig:cmb}
\end{figure}

Figure \ref{fig:cmb} shows the model predictions in the $(n_s,r)$ plane together with the joint Planck 2018 + BK18 constraints~\cite{BICEP:2021}. In the limit $\Omega=\xi^{2}$, satisfied by our benchmark, the slow-roll expressions admit the analytic approximation~\cite{Cecchini:2024}
\begin{equation}
3(1-n_s)^2-r\simeq\frac{64}{3}\xi^{2}\,.
\label{eq:nsranal}
\end{equation}
For fixed $\xi$, this defines a one-parameter curve in the $(n_s,r)$ plane, with $N_*$ determining the position along it. Its left-hand side measures the departure from pure Starobinsky inflation. Indeed, in the limit $\xi\to0$, the standard Starobinsky consistency relation $r\simeq3(1-n_s)^2$ is recovered, whereas a non-vanishing $\xi$ displaces the predictions from the pure $R^2$ curve by an amount proportional to $\xi^2$. This departure has a direct late-time counterpart. As shown in Sec.~\ref{sec:DE}, the same coupling $\xi$ fixes the slope of the Goldstone potential and, therefore, controls the present DE equation of state. Moreover, the condition $\Omega=\xi^2$ selects the Minkowski branch of the post-inflationary vacuum manifold, ensuring that the late-time DE sector has a purely unimodular origin, as discussed in Sec.~\ref{sec:lambda_naturalness}.

\subsection{Perturbative reheating}\label{sec:reheating}

Since the potential is quadratic near its minimum, the coherent oscillations of $\rho$ redshift as pressureless matter, with $\langle w_{\rho}\rangle=0$ \cite{Albrecht:1982b,Dolgov:1982,Turner:1983}; we therefore model reheating by treating $\rho$ as a pressureless fluid decaying perturbatively into a radiation bath. We parametrise this transition through a constant decay rate $\Gamma$, assuming the inflaton energy is efficiently transferred into the SM sector. For our benchmark, we adopt $\Gamma = 10^{-6} M_P$ \cite{Kofman:1994,Kofman:1997,Allahverdi:2010,Drewes:2014}, a phenomenological choice that implies direct non-gravitational couplings between the scalar sector and the SM. Under this assumption, the radiation bath is populated by the full set of $g_* = 106.75$ relativistic degrees of freedom of the SM. We assume the branching fraction to non-relativistic matter is negligible during this epoch. The background densities then obey
\begin{equation}
  \frac{d\rho_{\rho}}{dN}= -3\rho_{\rho} - \frac{\Gamma}{H}\,\rho_{\rho}\,,\qquad\quad  
  \frac{d\rho_{r}}{dN}  = -4\rho_{r} + \frac{\Gamma}{H}\,\rho_{\rho}\,,\qquad \quad \frac{d \rho_m}{dN} = -3 \rho_m \,,
  \label{eq:drhor}
\end{equation}
with $H^{2}=(\rho_{\rho}+\rho_{r}+\rho_{m})/(3M_P^{2})$. Reheating ends at $N_{\rm reh}$, defined as the first e-fold at which $\rho_{r}=\rho_{\rho}$, and the duration of reheating is $\Delta N_{\rm reh}=N_{\rm reh}-N_{\rm end}$.   The reheating temperature follows from the radiation density at
$N_{\rm reh}$,
\begin{equation}
  T_{\rm reh}= \left(\frac{30}{\pi^{2}\,g_{*}}\right)^{1/4}\,\rho_{r}(N_{\rm reh})^{1/4}.
\label{eq:Treh}
\end{equation}
For the benchmark parameters $\alpha=2.496\times10^{10}$, $\xi=0.01$, $\Omega=10^{-4}$, and $\Gamma=10^{-6}\,M_{P}$ we find $\rho_{r}(N_{\rm reh})\simeq 4.7\times10^{-13}\,M_{P}^{4}$, which yields
\begin{equation}
  T_{\rm reh} \simeq 8.3\times10^{14}~\text{GeV}\,.
\label{eq:Treh_num}
\end{equation}
Here $\rho_{r}(N_{\rm reh})$ is read directly from the numerical solution of the Boltzmann system; the thermal formula~\eqref{eq:Treh} then converts that energy density into a temperature. The chosen $\Gamma = 10^{-6}\,M_P$ satisfies $\Gamma \lesssim H_{\rm inf} \sim 5\times10^{-6}\,M_P$, placing the model in the standard perturbative reheating regime, in which reheating completes when $H \sim \Gamma$. In this regime, one expects $T_{\rm reh} \sim (\Gamma M_P)^{1/2} \sim 10^{-3}\,M_P \sim 10^{15}$~GeV, providing an analytic sanity check on the numerical result~\eqref{eq:Treh_num}.

The number of e-folds at CMB horizon crossing follows from tracing the pivot scale\footnote{We set $a_{0}=1$ throughout, so that $k_{*}$ denotes the physical pivot wavenumber today; in natural units $H_{*}$ and $k_{*}$ both carry mass dimension one, so the first term of~\eqref{eq:Nstar_reh} is dimensionless.} $k_{*}=0.05\,\mathrm{Mpc}^{-1}$ through reheating and radiation domination via entropy conservation \cite{Cook:2015,Dai:2014},
\begin{equation}
  N_{*} = \ln\frac{H_{*}}{k_{*}}  + \frac{1}{3}\ln\frac{\rho_{\rm reh}}{\rho_{\rm end}}
          + \frac{1}{3}\ln\frac{g_{*s,0}}{g_{*s,\rm reh}}   + \ln\frac{T_{0}}{T_{\rm reh}}\,,
\label{eq:Nstar_reh}
\end{equation}
with $T_{0}=2.73$~K, $g_{*s,0}=3.91$, $g_{*s,\rm reh}=106.75$, and $H_{*}=\sqrt{V_{*}/3\,M_P^2}\simeq5.45\times10^{-6}\,M_{P}$. Inserting our numerical results into \eqref{eq:Nstar_reh} yields $N_* \simeq 55$. The sensitivity of $N_*$ to the decay rate follows directly from~\eqref{eq:Nstar_reh}. Since reheating completes when $H\sim\Gamma$, one has $\rho_{\rm reh}\propto\Gamma^{2}$ and $T_{\rm reh}\propto\Gamma^{1/2}$, so that
\begin{equation}
  \frac{dN_*}{d\ln\Gamma}
  = \frac{1}{3}\,\frac{d\ln\rho_{\rm reh}}{d\ln\Gamma}- \frac{d\ln T_{\rm reh}}{d\ln\Gamma}
  = \frac{2}{3}-\frac{1}{2} = \frac{1}{6}\,,
\label{eq:dNdGamma}
\end{equation}
i.e.~$N_*$ shifts by $\\ln 1/6\simeq0.4$ e-folds per order of magnitude in $\Gamma$. The benchmark $\Gamma=10^{-6}M_P$ lies within a factor of a few of the perturbativity limit $\Gamma\lesssim H_{\rm inf}$, so the instantaneous-reheating limit raises $N_*$ by at most $\sim0.3$ e-folds, to $N_*\lesssim55.5$, while any slower reheating
lowers it. 

The corresponding CMB observables lie within the Planck+BK18 $2\sigma$ region. Note that $\alpha$ controls the overall height of the inflationary plateau and hence fixes the scalar amplitude $A_{s}\simeq2.1\times10^{-9}$, but leaves $n_{s}$, $r$, and $N_{*}$ essentially unchanged: a $22\%$ increase to $\alpha=3.056\times10^{10}$ shifts $N_{*}$ by less than $0.1$ e-folds. For reference, Fig.~\ref{fig:cmb} also shows $N_{*}=60$, which lies closer to the Planck $1\sigma$ boundary but is not a prediction of our reheating analysis.

\section{Hot Big Bang era}\label{sec:4}

After reheating, the Universe enters the standard thermal history \cite{Kolb:1990,Weinberg:2008cosmo}. We track three stages: radiation domination, during which the radiation energy density \(\rho_r\) redshifts from its value at reheating; Big Bang nucleosynthesis (BBN), during which we verify that the quintessence energy density satisfies the observational constraints in \cite{Cyburt:2015mya}; and matter
domination, throughout which the field remains cosmologically inert. 

\subsection{Radiation domination}\label{sec:BBN}

\begin{figure}
    \centering
    \includegraphics[width=0.95\linewidth]{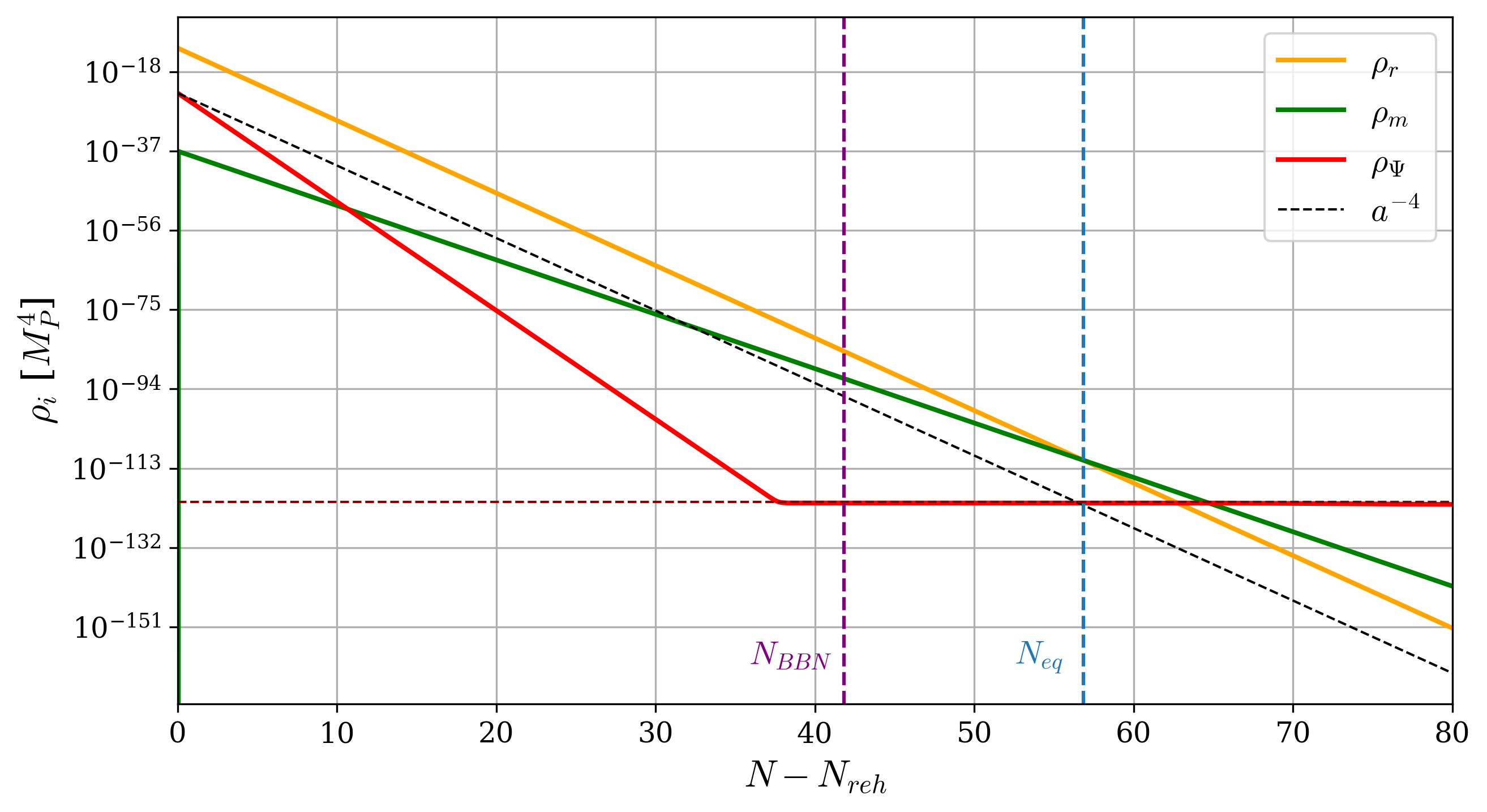}
    \caption{Evolution of the energy densities $\rho_i$ as a
function of e-folds $N-N_{\rm reh}$ since reheating. Orange: radiation $\rho_r$, decaying as $a^{-4}$ (black dashed reference line). Green: matter $\rho_m$. Red: quintessence $\rho_\Psi$, initialised with $d\Psi/dN=10^{-5}$ at the end
of reheating; the early steep decay (slope $a^{-6}$) is a transient kination phase, ending at $N-N_{\rm reh}\simeq38$ when the kinetic energy drops below $V_{\rm UG}$. The field then freezes at $\rho_\Psi\simeq V_{\rm DE}^{(0)}\simeq10^{-120}\,M_P^{4}$ (dark-red dashed line). Matter-radiation equality (navy dashed line) occurs at $N_{\rm eq} \simeq 57$, BBN (purple dashed line) at $N_{\rm BBN} \simeq 42$.}
    \label{fig: rho_full_ev}
\end{figure}

During radiation domination, the scale factor grows as $a\propto t^{1/2}$ and the radiation energy density redshifts as
\begin{equation}
  \rho_{r}(N) = \rho_{r}(N_{\rm reh})\,e^{-4(N-N_{\rm reh})}\,,
\label{eq:rhoredshift}
\end{equation}
where we have used $\rho_{r}\propto a^{-4}$. This behaviour is confirmed numerically in Fig.~\ref{fig: rho_full_ev}, where the orange curve follows the $a^{-4}$ reference line (black dashed) throughout the radiation-dominated phase. The Goldstone coordinate \(\chi\) remains close to zero after
inflation, apart from the residual decaying mode discussed in Sec.~\ref{sec:DE}. Consequently, the canonically normalised quintessence field
\(\Psi\equiv e^{b(\rho_{\min})}\chi\) does not participate in the thermal bath. After the transient early kination phase, its energy density is entirely potential and asymptotically constant (dark-red dashed line in Fig.~\ref{fig: rho_full_ev}).

BBN occurs at $T_{\rm BBN}\simeq 1$~MeV, when
$g_{*}(T_{\rm BBN})=10.75$ \cite{Cyburt:2015mya,Pitrou:2018,Cooke:2014}.  The radiation energy density at that epoch is
\begin{equation}
  \rho_{r}(N_{\rm BBN}) = \frac{\pi^{2}}{30}\,g_{*}(T_{\rm BBN})\,
  T_{\rm BBN}^{4} \simeq 1.01\times10^{-85}\,M_{P}^{4}\,.
\label{eq:rhoBBN}
\end{equation}
Matching $\rho_{r}(N_{\rm reh})\simeq 4.7\times10^{-13}\,M_{P}^{4}$ and \eqref{eq:rhoBBN} via \eqref{eq:rhoredshift} gives the number of e-folds elapsed between reheating and BBN,
\begin{equation}
  \Delta N_{\rm BBN} \;\equiv\; N_{\rm BBN} - N_{\rm reh}
  \;=\; \frac{1}{4}\ln\!\frac{\rho_{r}(N_{\rm reh})}{\rho_{r}(N_{\rm BBN})}
  \;\simeq\; 42\,,
\label{eq:DeltaN}
\end{equation}
in good agreement with the standard estimate of $\sim40$--45 e-folds from the end of reheating to BBN, as indicated by the purple dashed marker in Fig.~\ref{fig: rho_full_ev}.

The observational constraint on an additional scalar-field energy component at BBN can be expressed as $\Omega_{\Psi}(T_{\rm BBN})<0.045$ at 95\% C.L.~\cite{Bean:2001,Schramm:1998,Cyburt:2015mya}. In our model, any residual kinetic energy of $\Psi$ left over from reheating redshifts as $a^{-6}$ (an early kination phase) until it drops to $V_{\rm UG}\simeq V_{\rm DE}^{(0)}$. Denoting by $N_{\rm freeze}$ the e-fold at which this occurs, and by $\rho_{\Psi,0}$ the initial Goldstone energy density at the end of reheating, 
$N_{\rm freeze}-N_{\rm reh}= \tfrac16\ln(\rho_{\Psi,0}/V_{\rm DE}^{(0)})\lesssim\tfrac16\ln(\rho_{\rm tot,reh}/V_{\rm DE}^{(0)}) \simeq42\simeq\Delta N_{\rm BBN}$, i.e.~the field is frozen no later than BBN for any initial energy density not exceeding the total at reheating (and at $\simeq38$ for our initial condition, cf.~Sec.~\ref{sec:DE}). At BBN, the energy density of $\Psi$ is therefore entirely potential, $\rho_\Psi=V_{\rm DE}^{(0)}$, and its ratio to the radiation energy density is
\begin{equation}
  \Omega_{\Psi}(T_{\rm BBN}) \;=\;
  \frac{\rho_{\Psi}}{\rho_{r}(N_{\rm BBN})}
  \;\simeq\; 10^{-35}\,,
\label{eq:OmegaPsi}
\end{equation}
which is well below the observational bound.  This extreme suppression, visible in Fig.~\ref{fig: rho_full_ev} as the vast gap between the red and orange curves at $N_{\rm BBN}$, confirms that the thawing quintessence field is entirely irrelevant during BBN. The radiation-dominated epoch is then indistinguishable from that of $\Lambda$CDM at any epoch probed by BBN or earlier observations.

\begin{figure}
    \centering
    \includegraphics[width=0.95\linewidth]{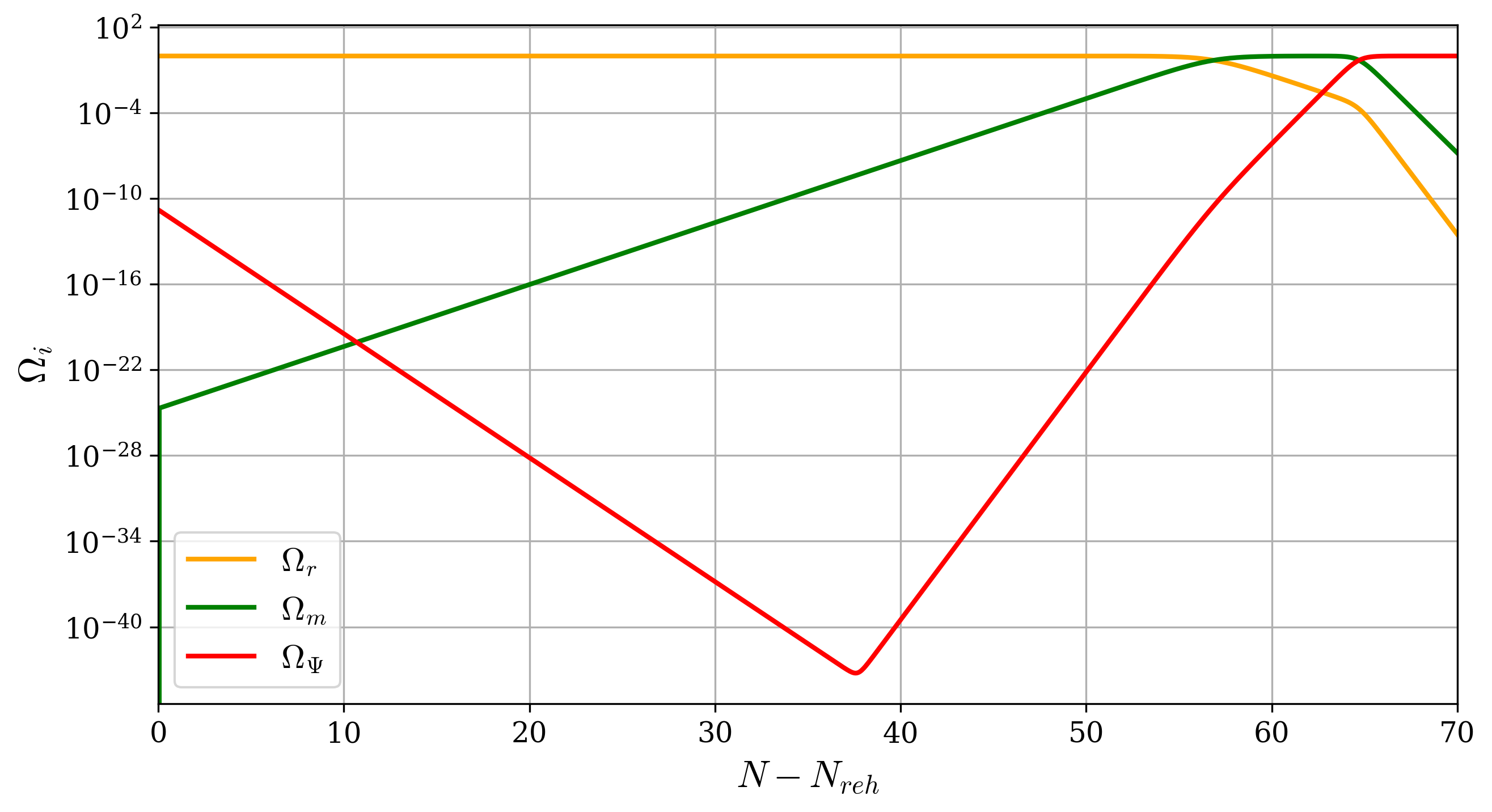}
    \caption{Evolution of the fractional energy densities $\Omega_i \equiv \rho_i / \rho_{\rm tot}$ as a function of e-folds $N - N_{\rm reh}$ since the end of reheating. In orange: radiation $\Omega_r$, in green: pressureless matter $\Omega_m$; in red: quintessence field $\Omega_\Psi$. Matter-radiation equality occurs at $N_{eq}\simeq 57$.}
    \label{fig: densities_par_ev}
\end{figure}

\subsection{Matter domination}
Matter--radiation equality occurs when \(\rho_m=\rho_r\). Combining \eqref{eq:rhoredshift} with \(\rho_m\propto a^{-3}\) gives
\begin{equation}
  N_{\rm eq} - N_{\rm reh}= \ln\!\frac{\rho_r(N_{\rm reh})}{\rho_m(N_{\rm reh})} \;\simeq\; 57\,,
\label{eq:Neq}
\end{equation}
consistent with the blue dashed marker in Fig.~\ref{fig: rho_full_ev}. Here, $\rho_m(N_{reh})$ is fixed by normalising the matter density to the observed $\Omega_m = 0.31$ today. Equality at $N_{eq} - N_{reh} \simeq 57$ is then a consistency check of the thermal history rather than an independent prediction.

After equality the Universe enters matter domination, $a \propto t^{2/3}$, and the matter energy density continues to redshift as \cite{Weinberg:2008cosmo}
\begin{equation}
  \rho_m(N) = \rho_m(N_{\rm eq})\,e^{-3(N - N_{\rm eq})}\,.
\label{eq:rhomatter}
\end{equation}

Throughout matter domination, \(m_\Psi\ll H\), so Hubble friction keeps \(\Psi\simeq0\) and \(\rho_\Psi\simeq V_{\rm UG}\sim10^{-120}M_P^4\), as shown in Fig.~\ref{fig: rho_full_ev}. The field begins to \emph{thaw}, i.e.~to roll away from $\Psi=0$, only when $H$ drops to the level of the effective mass
\begin{equation}
  m_\Psi  \simeq  4\gamma\sqrt{3\Omega_{\rm DE}\,} H_0\,.
\label{eq:mPsi}
\end{equation}
For our benchmark parameters $\gamma\simeq 0.058$ and $\Omega_{\rm DE}\simeq 0.7$, this condition is met near the present epoch $H_0\sim 3 \,m_\Psi$, signalling the onset of DE domination.  We discuss this transition and the resulting DE phenomenology in Sec.~\ref{sec:DE}.
Fig.~\ref{fig: densities_par_ev} confirms this behaviour: following matter-radiation equality at $N_{\rm eq}\simeq 57$,
$\Omega_m$ rises toward $\Omega_m\simeq 0.31$ while $\Omega_r$ decreases and $\Omega_\Psi$ remains negligible throughout, so the Universe evolves as a standard matter-dominated FLRW spacetime until DE becomes relevant near the present epoch.

\section{Late-time cosmology}\label{sec:DE}

On the Minkowski branch, the late-time dynamics are governed by the unimodular potential evaluated at \(\rho=\rho_{\min}\). The quintessence field then experiences the exponential potential \eqref{eq:Vexp}. This is the classic exponential quintessence potential, and it belongs to the class of \emph{thawing} DE models \cite{Ratra:1988,Wetterich:1995,Caldwell:1998,Carroll:2001,Peebles:2003,Frieman:2008,Tsujikawa:2013}.

The equation of motion for $\Psi$ in a flat FLRW background is
\begin{equation}
  \ddot\Psi + 3H\dot\Psi + V_{\rm UG}'(\Psi) = 0\,,
\label{eq:EOM_Psi}
\end{equation}
where primes denote $d/d\Psi$, and
\begin{equation}
  H^2 = \frac{1}{3M_P^2}\left(\rho_r + \rho_m + \rho_\Psi\right)\,,\qquad 
  \rho_\Psi = \frac{1}{2}\dot\Psi^2 + V_{\rm UG}(\Psi)\,.
\end{equation}
We integrate \eqref{eq:EOM_Psi} numerically from reheating to the present epoch (i.e.~$N_{\rm today}$), with initial conditions $\Psi=0$ and $d\Psi/dN=10^{-5}$ at the end of reheating, corresponding to $\rho_{\Psi,0}\simeq8\times10^{-24}\,M_P^{4}$. Any non-zero velocity of the Goldstone corresponds to the decaying $c_{2}$ mode of the scale-symmetry Noether charge~\eqref{eq:Ksol}, whose residual amplitude after inflation is model-dependent but strongly diluted. This velocity redshifts away during an early kination phase ($\rho_\Psi\propto a^{-6}$), after which the field freezes at $\rho_\Psi\simeq V_{\rm DE}^{(0)}$. The freeze-out epoch depends on this choice only logarithmically, $N_{\rm freeze}-N_{\rm reh}=\tfrac16\ln(\rho_{\Psi,0}/V_{\rm DE}^{(0)}) \simeq38$, and any smaller velocity freezes the field earlier (a vanishing one leaves it frozen throughout). It is bounded by $N_{\rm freeze}-N_{\rm reh}\lesssim\tfrac16\ln(\rho_{\rm tot,reh}/V_{\rm DE}^{(0)})\simeq42$, i.e.~no later than BBN, even in the extreme case where $\Psi$ carries the entire energy budget at reheating. The late-time observables are insensitive to this choice.

\subsection{Thawing dynamics}
\label{sec:5.1}

\begin{figure}
    \centering
    \includegraphics[width=0.95\linewidth]{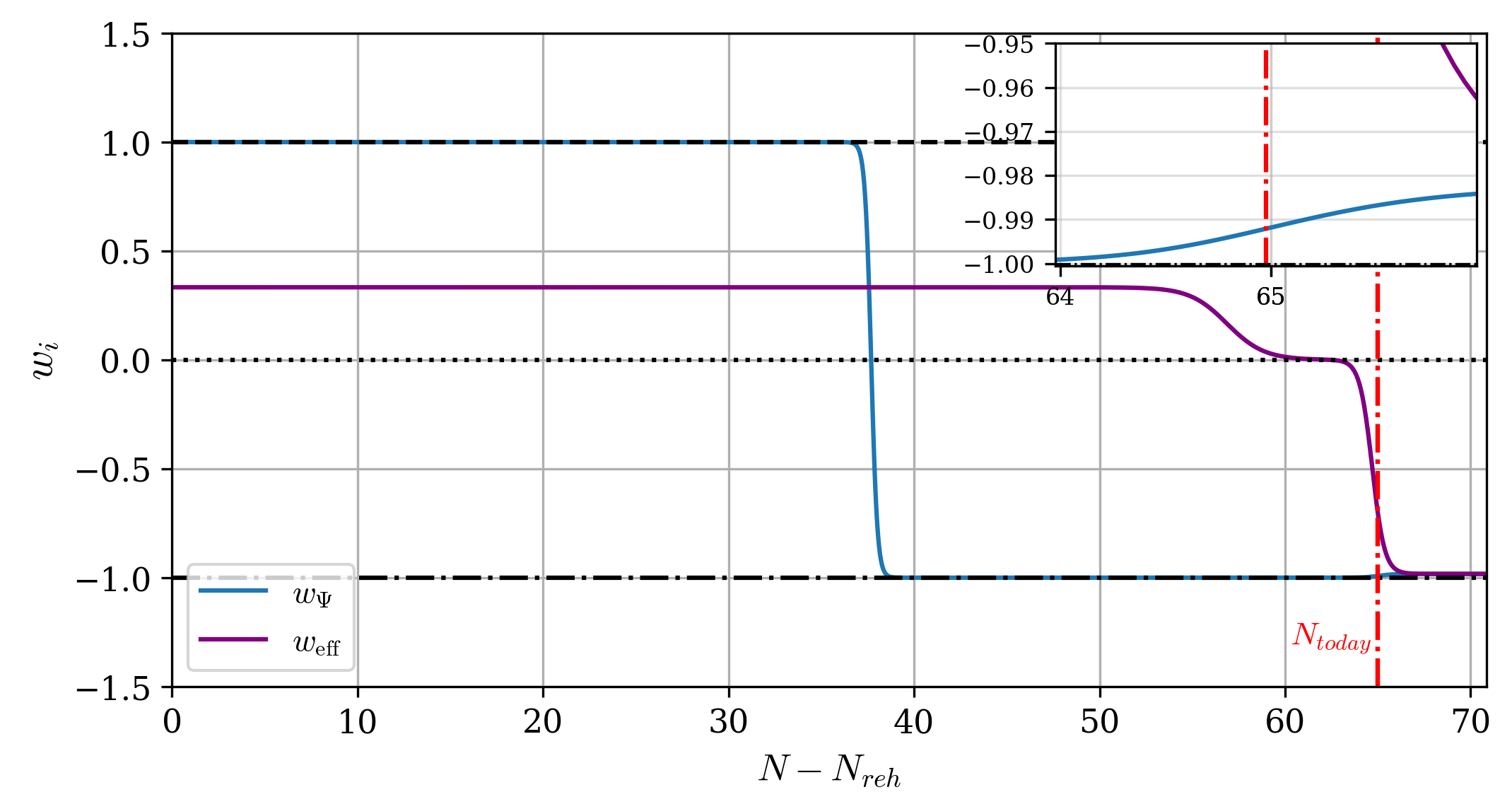}
    \caption{Evolution of the quintessence equation of state $w_\Psi$ (blue) and the total effective equation of state $w_{\rm eff}$ (purple) as functions of $N-N_{\rm reh}$. The horizontal lines mark $w=1$ (kination), $w=0$ (matter), and $w=-1$ (cosmological constant). The red dot-dashed vertical line indicates the present epoch $N_{\rm today}$. The inset shows a zoom around $N_{\rm today}$, revealing the onset of the thawing phase with $w_\Psi(N_{\rm today})\simeq-0.992$.}
    \label{fig: eq_of_state}
\end{figure}

Figure \ref{fig: eq_of_state} shows the evolution of the quintessence equation of state, $w_\Psi\equiv p_\Psi/\rho_\Psi$, together with the effective total equation of state, $w_{\rm eff}\equiv p_{\rm tot}/\rho_{\rm tot}$, from reheating to the present epoch. Immediately after reheating, the kinetic energy of $\Psi$ dominates over its potential, giving $w_\Psi\simeq1$. As the Universe expands, this contribution rapidly redshifts away and the potential takes over, causing $w_\Psi$ to fall sharply towards $-1$ at $N-N_{\rm reh}\simeq38$, in agreement with the analytic freeze-out estimate discussed above. The field then remains effectively frozen for most of the subsequent cosmological evolution.
The effective equation of state  follows the dominant background component $w_{\rm eff}\simeq1/3$ during radiation domination, approaches zero around matter--radiation equality at $N-N_{\rm reh}\simeq57$, and eventually becomes negative as DE starts to dominate. The inset of Fig.~\ref{fig: eq_of_state} shows that the quintessence field has only recently begun to thaw. At the present epoch, the numerical solution gives
\begin{equation}
  w_0 \;\equiv\; w_\Psi(N_{\rm today}) \;\simeq\; -0.992\,,
  \qquad
  w_a \;\equiv\; -\frac{dw_\Psi}{dN}\bigg|_{N_{\rm today}} \;\simeq\; -0.011\,.
\label{eq:w0wa}
\end{equation}
These definitions coincide with the standard CPL parametrisation, $w(a)=w_0+w_a(1-a)$~\cite{Chevallier:2001,Linder:2003}, so the pair $(w_0,w_a)$ can be compared directly with observational constraints in the corresponding parameter plane. The equation of state lies very close to, but slightly above, the cosmological-constant value $w=-1$, and drifts slowly upward as the field thaws, explaining the negative sign of $w_a$ in the convention~\eqref{eq:w0wa}.

At the present epoch, however, the field has not yet reached the asymptotic slow-roll attractor of an exponential potential,
\begin{equation}
  w_{\rm \Psi}^{\rm attractor} \simeq -1 +\frac{16\gamma^2}{3} \simeq -0.982\,,
\label{eq:wde_analytic}
\end{equation}
which is approached only once the field has settled into its late-time slow-roll motion~\cite{Chiba:2009,Scherrer:2008,Scherrer:2006,Dutta:2008qn,Linder:2008}. The numerical result therefore lies naturally between the frozen limit $w=-1$ and the asymptotic attractor, as expected for a field that has only recently entered the thawing regime.

\subsection{Early--late consistency relation}\label{sec:early-late}

The numerical value of $w_0$ in \eqref{eq:w0wa} can also be understood analytically. We work on the Minkowski branch $\Omega=\xi^2$ introduced in Sec.~\ref{sec:lambda_naturalness}, for which the scale-invariant sector leaves no residual vacuum energy and consider the small-$\xi$ regime. Since the field has only just begun to thaw, its energy density remains potential dominated, $\rho_\Psi\simeq V_{\rm DE}^{(0)}$, while the exponential potential~\eqref{eq:Vexp} can be linearised around its present value, $V_{,\Psi}\simeq-4\gamma\,V_{\rm DE}^{(0)}/M_P$. Radiation is already negligible at this stage, so the background can be approximated by a spatially flat matter-plus-effective-$\Lambda$ cosmology.

On the Minkowski branch, the minimum condition~\eqref{eq:vacuum_manifold} gives $\sinh^2u_{\rm min}=1/(2\xi)$ and hence $\cosh^2u_{\rm min}=(1+2\xi)/(2\xi)$. Substituting this into the canonical normalisation~\eqref{eq: psgold} and the definition~\eqref{eq:muPsi}, the exponential slope becomes
\begin{equation}
\gamma^2=\frac{1}{6\cosh^2 u_{\rm min}}=\frac{\xi}{3\,(1+2\xi)}\,,
\label{eq:muxi}
\end{equation}
which is exact on the Minkowski branch.

Under the approximations described above, the field equation~\eqref{eq:EOM_Psi} can be integrated once to give
\begin{equation}
\dot{\Psi}(t)=\frac{4\gamma V_{\rm DE}^{(0)}}{M_P\,a^3(t)}\int_0^{t}a^3(t')\,dt'\,.
\label{eq:Psisol}
\end{equation}
Using the exact flat matter-plus-$\Lambda$ solution~\cite{Weinberg:2008cosmo}, together with $V_{\rm DE}^{(0)}=3M_P^2H_0^2\Omega_{\rm DE}$ and $1+w_0\simeq\dot{\Psi}_0^2/V_0$, the time integral can be performed analytically, yielding
\begin{equation}
1+w_0=\frac{16\gamma^2}{3}F(\Omega_{\rm DE})\,,\qquad
F(\Omega_{\rm DE})=\frac{1}{\Omega_{\rm DE}}\left[1-\frac{1-\Omega_{\rm DE}}{\sqrt{\Omega_{\rm DE}}}\,\mathrm{arctanh}\sqrt{\Omega_{\rm DE}}
\right]^{2}.
\label{eq:w0kappa}
\end{equation}
For the observed DE density fraction $\Omega_{\rm DE}=0.69$, one finds $F=0.448$. For the benchmark $\xi=10^{-2}$, this gives $w_0\simeq-0.992$, in excellent agreement with the numerical result~\eqref{eq:w0wa}. 
The fact that \(F(0.69)=0.45<F(1)=1\) is consistent with the statement in Refs.~\cite{Casas:2017wjh,Casas:2018fum} that \(F\) interpolates from 0 to 1. A particularly transparent consequence of \eqref{eq:w0kappa} is that the non-minimal coupling (or $\gamma$) can be eliminated entirely in favour of observable quantities. Combining it with the inflationary prediction~\eqref{eq:nsranal}, we obtain
\begin{equation}
(1-n_s)^2-\frac{r}{3}\simeq\left(\frac{3(1+w_0)}{2F(\Omega_{\rm DE})}\right)^2\,,\qquad \qquad (\textrm{SI+UG $R^2$ gravity})\,. 
\label{eq:early_late_consistency}
\end{equation}
This relation shows that the deformation away from the pure Starobinsky relation and the present thawing signal are controlled by the same non-minimal coupling $\xi$, removing the freedom to tune the late-time dynamics independently of inflation, as shown in Fig.\,\ref{fig:ns_w0_plane} for $N_*=55$, $\xi=10^{-2}$ and $\Omega=10^{-4}$.

It is instructive to compare \eqref{eq:early_late_consistency} with the consistency relations of Higgs--Dilaton cosmology \cite{Garcia-Bellido:2011kqb,Casas:2018fum}. Eliminating the number of inflationary e-folds between the Higgs--Dilaton expressions for \(n_s\) and \(r\), and specialising to the universal regime \(|\kappa_c|\simeq1/6\), the associated consistency relation coincides,  at leading order, with \eqref{eq:early_late_consistency}.  Indeed, once the different normalisation of the non-minimal coupling in the two actions is accounted for, which amounts to setting \(\xi = 3\xi_\chi\) with $\xi_{\chi}$ the dilaton coupling in \cite{Garcia-Bellido:2011kqb,Casas:2018fum}, the slope \eqref{eq:muxi} reproduces the Higgs--Dilaton expression \(\gamma^2=\xi_\chi/(1+6\xi_\chi)\) exactly. 
The two constructions therefore share not only the general mechanism connecting inflation and thawing DE but also the same observable consistency relation in their common universal regime. Their distinction lies instead in the microscopic origin of the inflationary sector and in corrections away from this limit.

The consistency relation~\eqref{eq:early_late_consistency} relies on the absence of a residual vacuum energy in the scale-invariant sector. For $\lambda>0$, after the radial field has relaxed to its minimum, the effective late-time potential becomes
\begin{equation}
    U(\Psi)=V(\rho_{\min}) +V_{\rm DE}^{(0)}e^{-4\gamma\Psi/M_P},
    \qquad   
V(\rho_{\min})=\frac{9M_P^4}{4}\frac{\lambda}{\Omega},
    \label{eq:DE_constant_exponential}
\end{equation}
with $V_{\rm DE}^{(0)}>0$ the exponential normalisation introduced in \eqref{eq:Vexp}. In the presence of a residual constant $V(\rho_{\min})$, the observed dark-energy density fixes the sum of the two potential contributions, up to a small kinetic correction, rather than the exponential
contribution alone. The matching prescription of Eq.~\eqref{Lambda_obs}
must therefore be modified accordingly. The slope $\gamma$ is still given by the general expression~\eqref{eq:muPsi} but the specialised expression~\eqref{eq:muxi} holds only on the Minkowski branch. 

To quantify the effect of the constant, we define the present exponential fraction of the potential energy,
\begin{equation}
    \mathcal E_0\equiv \frac{V_{\rm DE}^{(0)}e^{-4\gamma\Psi_0/M_P}}
    {U(\Psi_0)}, \qquad 0\leq\mathcal E_0\leq1,
    \label{eq:DE_exponential_fraction}
\end{equation}
where $\Psi_0$ is the present field value. The total dark-energy density and pressure are
\begin{equation}
    \rho_{\rm DE}=\frac12\dot\Psi^2+U(\Psi),
    \qquad
    p_{\rm DE}=\frac12\dot\Psi^2-U(\Psi).
    \label{eq:DE_total_density_pressure}
\end{equation}
We denote their present ratio by $w_{{\rm DE},0}\equiv p_{{\rm DE},0}/\rho_{{\rm DE},0}$, which reduces to $w_0$ as defined in Eq.~\eqref{eq:w0wa} on the Minkowski branch. Within the same potential-dominated thawing approximation used
in deriving Eq.~\eqref{eq:w0kappa},  the field displacement is small and the
potential gradient can be approximated by
\begin{equation}
    U_{,\Psi}\simeq -\frac{4\gamma}{M_P}\mathcal E_0 U(\Psi_0).
    \label{eq:DE_total_gradient}
\end{equation}
Repeating the integration in Eq.~\eqref{eq:Psisol}, with $U(\Psi_0)\simeq3M_P^2H_0^2\Omega_{\rm DE}$, gives
\begin{equation}
    1+w_{{\rm DE},0}  \simeq \frac{16\gamma^2}{3} F(\Omega_{\rm DE})\mathcal E_0^2.
    \label{eq:thawing_with_constant}
\end{equation} 
Thus, at fixed $\gamma$ and total dark-energy density, the departure from $w=-1$ is suppressed by the square of the exponential fraction \eqref{eq:DE_exponential_fraction}. The pure-exponential result is recovered for $\mathcal E_0=1$, while the limit $\mathcal E_0\to0$ approaches $\Lambda$CDM. 

A residual constant of order the observed dark-energy density is negligible compared with the inflationary potential scale.
Indeed, its ratio to $9M_P^4/(4\alpha)$ is
$\alpha\lambda/\Omega$, so that such a small residual constant
implies $\Omega\simeq\xi^2$. Its effects on the inflationary predictions and on $\gamma$ can therefore be neglected in the parameter region considered here.
Under this additional approximation, and retaining the small-$\xi$ and thawing approximations underlying Eq.~\eqref{eq:early_late_consistency}, the early--late
relation becomes
\begin{equation}
    1+w_{{\rm DE},0}
    \simeq
    \mathcal E_0^2
    \frac{2F(\Omega_{\rm DE})}{3}
    \sqrt{(1-n_s)^2-\frac{r}{3}}.
    \label{eq:consistency_with_constant}
\end{equation}
Consequently, to this accuracy,
\begin{equation}
    0\leq1+w_{{\rm DE},0}
    \lesssim
    \frac{2F(\Omega_{\rm DE})}{3}
    \sqrt{(1-n_s)^2-\frac{r}{3}},
    \label{eq:consistency_bound}
\end{equation}
with the upper envelope attained on the Minkowski branch. The equality of Eq.~\eqref{eq:early_late_consistency} is therefore replaced by an approximate upper bound on the present thawing behaviour.

The suppression derived above concerns the evolution through the present epoch. For any strictly positive residual constant, the subsequent evolution differs qualitatively from that of the pure exponential: as the field rolls towards larger $\Psi$, the exponential contribution decreases and the constant eventually dominates, leading to asymptotic de Sitter expansion with $w_{\rm DE}\to -1$. Thus, present-day thawing
can be followed by a return towards $w_{\rm DE}=-1$ in the future. This discussion assumes nonnegative residual vacuum energy and
a positive exponential contribution.

\subsection{Confrontation with observations}
\label{sec:datasets}
We now return to the Minkowski branch, $\lambda=0$, for the
comparison with observations. A model-independent diagnostic of the thawing regime is the Caldwell--Linder bound~\cite{Caldwell:2005},
\begin{equation}
  -3(1+w_0)\leq w_a\leq-(1+w_0)\,.
\label{eq:CL}
\end{equation}
The values obtained in \eqref{eq:w0wa} satisfy this condition, confirming that the pseudo-Goldstone field is currently in the thawing regime.

\begin{figure}
    \centering
    \includegraphics[width=0.8\linewidth]{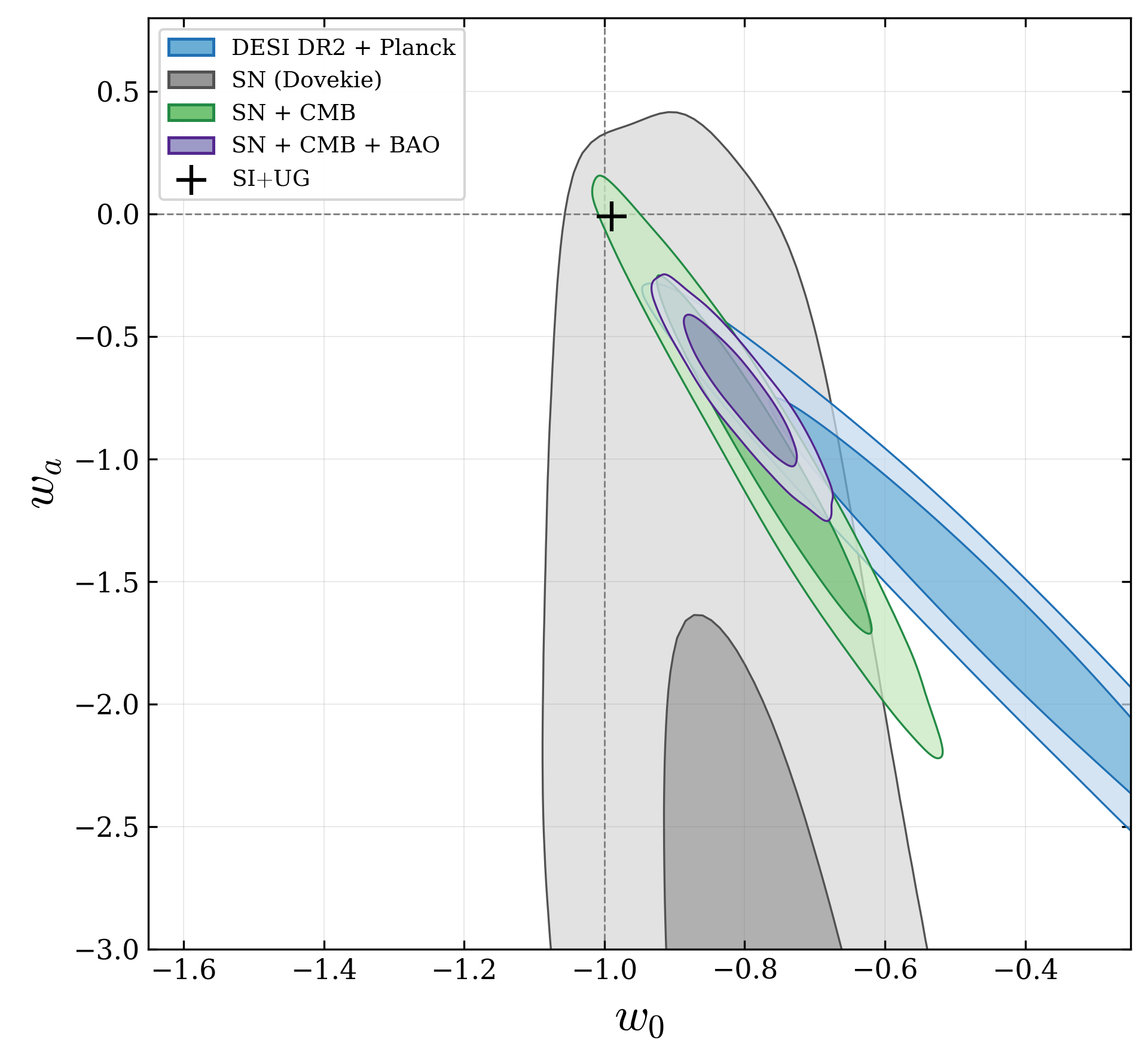}
    \caption{Combined $(w_0,w_a)$ constraints from Ref.~\cite{Popovic:2025}. Blue: DESI DR2 + Planck+ACT+SPT; grey: DES-Dovekie (wide prior); green: DES-Dovekie + Planck+ACT+SPT; purple: DES-Dovekie + Planck+ACT+SPT + DESI DR2. The star marks the SI$+$UG prediction $(w_0,w_a)=(-0.992,-0.011)$.}
    \label{fig: desi_dovekie}
\end{figure}

Figure~\ref{fig: desi_dovekie} compares the SI$+$UG prediction with the marginalised posteriors obtained in Ref.~\cite{Popovic:2025} using the recalibrated DES-Dovekie supernova sample. The prediction lies very close to the $\Lambda$CDM point and falls within the $95\%$ region of the $(w_0,w_a)$ plane for the SN$+$CMB combination, which exhibits the weakest preference for evolving DE among the dataset combinations considered. The addition of DESI~DR2 BAO shifts the preferred region towards larger departures from $w=-1$, placing both $\Lambda$CDM and the SI$+$UG prediction outside the corresponding $68\%$ contour. This behaviour is consistent with the observation that the current evidence for dynamical DE is driven primarily by the BAO measurements: removing DESI substantially reduces the preference for $w\neq-1$~\cite{Elbers:2025vlz,Popovic:2025}. The official DESI~DR2 analysis employing the Pantheon$+$ supernova sample~\cite{Brout:2022vxf} finds an even stronger preference for evolving DE, illustrating the sensitivity of the result to the adopted supernova calibration~\cite{Elbers:2025vlz}.

The statistical interpretation of these results remains less settled than their quoted significances alone might suggest. Using the recalibrated DES-Dovekie sample, Ref.~\cite{Popovic:2025} reduces the apparent tension with $\Lambda$CDM from $4.2\sigma$ to $3.2\sigma$, while the corresponding Bayesian odds provide only weak support for the $w_0w_a$CDM extension. A direct Bayesian evidence calculation reaches a similar conclusion, showing that combinations displaying an apparently significant frequentist preference for evolving DE need not favour the extended model once the parameter-volume penalty and the tension between datasets are properly taken into account~\cite{Bayes:2025}. We therefore regard the present observational situation as suggestive, but not yet conclusive.

For the SI$+$UG framework, however, the practical impact of this debate is presently limited. Its predicted departure from a cosmological constant, $|1+w_0|\simeq8\times10^{-3}$, is approximately seven times smaller than the current observational uncertainty, $\sigma(w_0)=0.054$~\cite{Popovic:2025}, placing the prediction only $0.15\sigma$ from the $\Lambda$CDM point. Current late-time observations therefore do not meaningfully distinguish the SI$+$UG prediction from a cosmological constant. Consequently, whether the present BAO-driven preference reflects new physics or residual systematics has little impact on the status of the model at present. Resolving this question will require future measurements from DESI~DR3, Euclid, and the Rubin Observatory~\cite{Amendola:2016saw,Rubin:2021}.

The more distinctive prediction of the framework is instead the correlation between the inflationary and DE sectors. Unlike phenomenological reconstructions of $w(z)$ or alternative-gravity scenarios whose late-time dynamics are adjusted to reproduce the data~\cite{Ye:2025}, the SI$+$UG framework predicts the DE equation of state from the inflationary attractor. A single pair $(\xi,\Omega)$ simultaneously determines $n_s$, $r$, $w_0$, and $w_a$, so the two cosmological epochs cannot be tuned independently.

The observational comparisons presented here should nevertheless not be interpreted as a fully joint test of the model. The $(n_s,r)$ constraints discussed in Sec.~\ref{subsec:infl} are derived assuming a cosmological constant, whereas the $(w_0,w_a)$ constraints above are obtained with the tensor-to-scalar ratio fixed to zero. Since the predicted values, $r\simeq2.9\times10^{-3}$ and $(w_0,w_a)\simeq(-0.992,-0.011)$, lie very close to the assumptions adopted in the respective analyses, we expect a combined $w_0w_a{\rm CDM}+r$ fit to induce only minor shifts in the inferred constraints. This expectation is further supported by the dedicated analysis of the Higgs–Dilaton model, which features an analogous early–late consistency relation \cite{Casas:2018fum}.

\section{Conclusions}\label{sec:6}

We have presented a minimal cosmological framework in which SI and UG establish a predictive connection between inflation and late-time cosmic acceleration. The central observation is that the Goldstone boson of spontaneously broken scale symmetry plays two qualitatively different roles throughout the cosmological evolution. During inflation, conservation of the scale-symmetry Noether current freezes this degree of freedom and renders the dynamics effectively single-field. At late times, the unimodular integration constant lifts the same flat direction and turns the Goldstone boson into a thawing DE field. The late-time sector is, therefore, not introduced independently, but emerges from the same symmetry structure that governs inflation.

A key result of the construction is that the slope of the Goldstone potential is fixed by the field-space geometry at the inflationary attractor rather than being a free
quintessence parameter.\footnote{This illustrates a more general connection between field-space geometry and inflationary observables recently emphasized in Ref.~\cite{Karananas:2023bog}.} This geometry excludes the matter-era tracker branch throughout the parameter space and places the entire inflationary viability region in the accelerating thawing regime. For our representative benchmark, the Goldstone field remains frozen until close to the present epoch and yields $
(w_0,w_a)\simeq(-0.992,-0.011)$. The predicted thawing signal is therefore small, but it is inherited from the inflationary sector rather than generated by an independently chosen DE potential.

The common geometric origin of inflation and DE leads to the appealing early--late consistency relation \eqref{eq:intro_early_late_relation}, which encapsulates the main result of this work. The same deformation that moves the inflationary predictions away from the pure $R^2$ relation generates the late-time thawing signal. This provides a simple observational target: improved measurements of $n_s$, $r$, and the DE equation of state directly test whether the departures from Starobinsky inflation and from $\Lambda$CDM share the common geometric origin predicted here.

The same rigidity that gives the model its predictive power also constitutes its main limitation. The non-minimal coupling $\xi$ controls both the inflationary spectral tilt and the late-time thawing amplitude, so the two epochs cannot be adjusted independently. Lowering this coupling raises $n_s$ only up to the pure Starobinsky limit while simultaneously driving $w_0\to-1$. Within the minimal Minkowski branch, an observable departure from $\Lambda$CDM and a spectral tilt compatible with the post-ACT preference therefore cannot be obtained simultaneously. Current post-ACT measurements consequently challenge not only the inflationary prediction, but the rigid early--late connection itself. Any successful extension must raise $n_s$ while preserving the geometric mechanism that fixes the Goldstone potential, rather than simply introducing additional freedom in the DE sector.

Several questions remain open. One is whether genuinely scale-invariant higher-curvature operators, such as combinations of the form $R^3/\phi^2$ or $R^4/\phi^4$, can shift the inflationary predictions without spoiling the scale-current attractor, the flat Goldstone direction, or the Minkowski vacuum structure. A second concerns the radiative stability of the branch $\Omega=\xi^2$, equivalently $\lambda=0$, which ensures that the scale-invariant sector leaves no residual vacuum energy and that the unimodular integration constant is the sole source of late-time acceleration. Since SI allows the operator $\lambda\phi^4$, quantum corrections may displace the theory from this branch. It will therefore be important to determine whether additional ultraviolet structures, such as asymptotic safety or supersymmetry, can protect the required relation and preserve the early--late consistency condition. A more complete treatment of reheating, including non-perturbative particle production and explicit decay channels into the SM particles \cite{Garcia-Bellido:2008ycs,Repond:2016sol,    Rubio:2019ypq,Piani:2023aof,Piani:2025dpy,Rubio:2026acm,Barman:2025lvk} and dark matter \cite{Bernal:2020qyu,Barman:2023opy}, would also sharpen the prediction for the thermal history.

Regardless of how these questions are ultimately resolved, the present work identifies a simple mechanism by which a single Goldstone degree of freedom connects two epochs of cosmic acceleration across the entire post-inflationary history. More importantly, this connection is quantitative rather than merely qualitative: the same field-space geometry predicts both the departure from pure Starobinsky inflation and the departure of DE from a cosmological constant. In this sense, the model does not simply place inflation and DE within a common framework; it identifies them as two observable manifestations of the same underlying scale-invariant geometry.

\acknowledgments

MDA thanks Massimiliano Rinaldi for drawing our attention to the analytic approximation discussed in Appendix \ref{app:lambda0}. MDA is supported by the Leverhulme Trust. J.~R. is supported by a Ram\'on y Cajal contract of the Spanish Ministry of Science and Innovation with Ref.~RYC2020-028870-I. This research was further supported by the project PID2022-139841NB-I00 of MICIU/AEI/10.13039/501100011033 and FEDER, UE, and the 2025 Leonardo Grant for Scientific Research and Cultural Creation from the BBVA Foundation. The BBVA Foundation is not responsible for the opinions, comments, and content included in the project and/or its resulting outcomes, which are the sole and exclusive responsibility of the authors.

\appendix

\section{The post-inflationary attractor}\label{sec:fixedpoint}

The two-field system (whose equations of motion in an FLRW background are given in Eqs.~\eqref{eq:EOM_rho}--\eqref{eq:friedmann}) is conveniently analysed as an autonomous dynamical system. Here, the dynamics are governed by the scale-invariant potential \eqref{eq:Vrho}, and we define the dimensionless phase-space variables as
\begin{equation}
  x \equiv \frac{\dot{\rho}}{\sqrt{6}\,H M_{P}}\,,
  \qquad
  y \equiv \frac{\sqrt{V}}{\sqrt{3}\,H M_{P}}\,,
  \qquad
  \Omega_{\chi} \equiv \frac{e^{2b(\rho)}\dot{\chi}^{2}}{6H^{2}M_{P}^{2}}\,,
\label{eq:autovars}
\end{equation}
which satisfy the Friedmann constraint $x^{2}+y^{2}+\Omega_{\chi}=1$.  We also introduce the logarithmic slope of the potential and the field-space curvature coupling,
\begin{equation}
  \lambda(\rho) \equiv M_{P}\frac{V_{,\rho}}{V}\,,
  \qquad\quad 
  \beta(\rho) \equiv M_{P}\,b_{,\rho}(\rho)\,.
\label{eq:lambdabeta}
\end{equation}
Moreover, the \(\chi\)-dependent source term can be rewritten as
$\frac{e^{2b(\rho)}b_{,\rho}\dot{\chi}^2}{\sqrt{6}H^2M_P}=\sqrt{6}\beta\Omega_{\chi}$. The equations of motion then yield the autonomous system
\begin{align}
  \frac{dx}{dN} &= -3x - \frac{\sqrt{6}}{2}\,\lambda\,y^{2}
    + \sqrt{6}\,\beta\,\Omega_{\chi} + 3x(x^{2}+\Omega_{\chi})\,,
  \label{eq:dxdN}\\
  \frac{dy}{dN} &= \frac{\sqrt{6}}{2}\,\lambda\,xy
    + 3y(x^{2}+\Omega_{\chi})\,,
  \label{eq:dydN}\\
  \frac{d\Omega_{\chi}}{dN} &=
    2\Omega_{\chi}\!\left(-3 - \sqrt{6}\,\beta\,x
    + 3x^{2} + 3\Omega_{\chi}\right)\,,
  \label{eq:dOcdN}
\end{align}
where $N=\ln a$. Since $\lambda$ and $\beta$ depend on $\rho$, the system is closed by the first of \eqref{eq:autovars}, namely
\begin{equation}
  \frac{d\rho}{dN} \;=\; \sqrt{6}\,M_{P}\,x \,,
\label{eq:drhodN}
\end{equation}
which makes it autonomous in the four variables $(x,y,\Omega_{\chi},\rho)$. One may verify that \eqref{eq:dxdN}--\eqref{eq:dOcdN} preserve the Friedmann constraint, i.e.~$d(x^{2}+y^{2}+\Omega_{\chi})/dN \equiv 0$. The system admits a fixed point at $(0,1,0,\rho_{\rm min})$: with $x=0$, \eqref{eq:drhodN} is automatically stationary, and \eqref{eq:dxdN} reduces the fixed-point condition to
\begin{equation}
  \lambda(\rho_{\rm min}) = 0
  \qquad\Longleftrightarrow\qquad
  V_{,\rho}(\rho_{\rm min}) = 0\,.
\label{eq:FPcond}
\end{equation}
Both the de Sitter saddle at $\rho=0$ and the minimum satisfy this condition; stability selects the latter. Linearising around the fixed point (with $\delta y$ eliminated through the
Friedmann constraint), the $\Omega_{\chi}$ direction decouples
with eigenvalue $-6$, while the $(x,\delta\rho)$ pair behaves as a damped oscillator with eigenvalues
\begin{equation}
  \mu_{\pm} \;=\; -\frac{3}{2}
  \pm\sqrt{\,\frac{9}{4}-\frac{3M_P^{2}\,V_{,\rho\rho}(\rho_{\rm min})}
  {V(\rho_{\rm min})}\,}\,,
\label{eq:FPeigen}
\end{equation}
whose real parts are negative for $V_{,\rho\rho}(\rho_{\rm min})>0$. The minimum therefore defines a stable attractor: the field is dynamically driven towards the minimum and trapped there. We denote the stable point by the subscript ``min''.
We stress that the fixed point exists as stated only for
$V(\rho_{\rm min})>0$, i.e.~for $\Omega>\xi^{2}$. On the benchmark boundary $\Omega=\xi^{2}$ the potential has a quadratic zero at $\rho_{\rm min}$ and the variable $y$ degenerates (see \eqref{eq:app-square}). The trapping mechanism, however, survives this limit unchanged. Indeed, for $V_{,\rho\rho}(\rho_{\rm min})>0$ the field has underdamped oscillations around $\rho_{\rm min}$, with $m_{\rho}\gg H$ after the end of inflation, whose amplitude is redshifted by Hubble friction. The associated energy density, which redshifts as pressureless matter, is subsequently transferred to radiation through the perturbative decay described in Sec.~\ref{sec:reheating}. This mechanism effectively ``locks'' the radial field, allowing the remaining scalar degree of freedom to dictate the late-time accelerated expansion of the Universe.

\section{Essentials of unimodular gravity}\label{app:ug}

UG is a restricted version of general relativity in which the metric determinant is held fixed~\cite{Anderson:1971,Buchmuller:1988a,
Buchmuller:1988b,Unruh:1989,Henneaux:1989}. The constraint
\begin{equation}
  \sqrt{-g} = \varepsilon_{0} = \mathrm{const}\,,
\label{eq:UGconstraint}
\end{equation}
can be imposed either as a gauge condition, reducing the diffeomorphism group Diff to its subgroup of transverse (volume-preserving) diffeomorphisms \(\mathrm{TDiff}\)~\cite{Buchmuller:1988a,Buchmuller:1988b,Alvarez:2005,Eichhorn:2015}, or through a Lagrange multiplier $\Lambda(x)$ added to a general action with matter fields~\cite{Unruh:1989,Henneaux:1989}, including possible higher-curvature contributions such as an $R^2$ term,
\begin{equation}
S = \int d^4 x \left[ \sqrt{-g}\,\mathcal{L}(g_{\mu\nu}, R, R^2, \Phi, \partial \Phi) + \Lambda(x)\left(\sqrt{-g} - 1\right) \right]\,.
\end{equation}
We set $\varepsilon_0=1$ without loss of generality. Varying the action with respect to $\Lambda(x)$ yields $\sqrt{-g} = 1$, while variation with respect to $g^{\mu\nu}$ gives the modified gravitational equations
\begin{equation}
\frac{\delta \left( \sqrt{-g}\,\mathcal{L} \right)}{\delta g^{\mu\nu}} - \frac{1}{2}\Lambda(x)\,\sqrt{-g} g_{\mu\nu} = 0\,.
\end{equation}
These can be rewritten in terms of the energy-momentum tensor $T_{\mu\nu}$ and an effective Einstein tensor $G_{\mu\nu}^{\mathrm{eff}}$ (including the contribution from the $R^2$ term or its scalar--tensor representation) as
\begin{equation}
G_{\mu\nu}^{\mathrm{eff}} = \frac{1}{M_P^2}\left( T_{\mu\nu} - \Lambda(x) g_{\mu\nu}\right)\,.
\end{equation}
Variation with respect to the matter fields $\Phi$ gives their standard equations of motion. The constraint term $\Lambda(x)(\sqrt{-g}-1)$ breaks the manifest diffeomorphism invariance of the full action, since the constant piece is not accompanied by the invariant volume density and therefore does not transform as a scalar density under general diffeomorphisms. The part proportional to $\sqrt{-g}\,\mathcal{L}$, however, retains full diffeomorphism invariance even in the presence of higher-curvature terms such as $R^2$, which preserve diffeomorphism invariance. Exploiting this symmetry and using the equations of motion, one obtains the identity
\begin{equation}
\int d^4 x \, \sqrt{-g}\, (\partial_\mu \Lambda)\, \zeta^\mu = 0\,,
\end{equation}
for arbitrary vector fields $\zeta^\mu(x)$, which implies $\partial_\mu \Lambda(x) = 0$. Equivalently, taking the covariant divergence of the field equations and using the generalized Bianchi identity together with $\nabla^\mu T_{\mu\nu} = 0$, one finds
\begin{equation}
\nabla_\nu \Lambda(x) = 0\,.
\label{eq: lambda_const}
\end{equation}
Therefore $\Lambda(x) = \Lambda_0$ is a constant of motion fixed by initial conditions. The resulting equations reduce to
\begin{equation}
G_{\mu\nu}^{\mathrm{eff}} = \frac{1}{M_P^2}\left( T_{\mu\nu} - \Lambda_0\, g_{\mu\nu}\right)\,,
\end{equation}
showing that, even in the presence of higher-curvature terms such as $R^2$, UG is classically equivalent to a fully diffeomorphism-invariant theory in which the cosmological constant arises as an integration constant rather than a parameter of the Lagrangian.

\section{Regularity of the Minkowski branch}\label{app:lambda0}

The benchmark condition $\Omega=\xi^{2}$, equivalent to $\lambda=0$, plays a central role in our construction as it makes the inflationary potential vanish at its minimum, so that the unimodular integration constant $\Lambda_{0}$ is the sole source of late-time DE. In the analysis of the same inflationary sector in \cite{Ghoshal:2022qxk}, however, the parameter space is restricted to
\begin{equation}
  \xi^{2} \;<\; \Omega \;\leq\; \frac{2}{\sqrt{3}}\,\xi^{2}
  \;\approx\; 1.1547\,\xi^{2}\,,
  \label{eq:app-window}
\end{equation}
where the upper bound is a genuine reality condition on the end of inflation (see below), while the lower bound follows from assuming $\lambda>0$ strictly. Our benchmark therefore lies on a boundary excluded in that analysis by construction rather than by dynamics. In this appendix, we show that the boundary is regular and that the condition $\epsilon_{V}=1$ admits a well-defined physical solution at $\Omega=\xi^{2}$, i.e.~inflation ends before the field reaches the minimum of the potential. We stress that the large-field expansion employed below is used only to make contact with the analytic treatment of \cite{Ghoshal:2022qxk}, in which the boundary was not taken into account.

In the large-field expansion adopted in \cite{Ghoshal:2022qxk}, $\rho \gg M_P$, one has $\sinh u \simeq \tfrac{1}{2}\,e^{u}$ with $u \equiv \rho/(\sqrt{6}\,M_P)$, and it is convenient to introduce
\begin{equation}
  x \;\equiv\; \exp\!\left(\sqrt{\tfrac{2}{3}}\,\frac{\rho}{M_P}\right)\,,\qquad x \gg 1 \,,
  \label{eq:app-x}
\end{equation}
so that $\sinh^{2}u \simeq x/4$ and 
\begin{equation}
  V(x) \;\simeq \frac{9M_P^{4}}{4\alpha} \left[\,1-\xi x+\frac{\Omega}{4}\,x^{2}\right]\,.
  \label{eq:app-Vlargefield}
\end{equation}
The first potential slow-roll parameter then reads
\begin{equation}
  \epsilon_{V}(x)
  = \frac{4x^{2}\left(\Omega x-2\xi\right)^{2}}
             {3\left(\Omega x^{2}-4\xi x+4\right)^{2}} \,,
  \label{eq:app-eps}
\end{equation}
and the end-of-inflation condition $\epsilon_{V}=1$ is a quartic equation in $x$, whose four roots \cite{Ghoshal:2022qxk}
\begin{align}
  x_{1}^{(\pm)} &= \frac{2(\sqrt{3}-1)\,\xi}{\Omega}
  \pm \frac{2}{3\Omega}\sqrt{3\,(2\sqrt{3}-3)\,(2\sqrt{3}\,\xi^{2}-3\Omega)}\,,
  \label{eq:app-x1}\\
  x_{2}^{(\pm)} &= -\frac{2(\sqrt{3}+1)\,\xi}{\Omega}
  \pm \frac{2}{3\Omega}\sqrt{3\,(2\sqrt{3}+3)\,(2\sqrt{3}\,\xi^{2}+3\Omega)}\,.
  \label{eq:app-x2}
\end{align}
The roots \eqref{eq:app-x2} are real for all $\xi,\Omega>0$ and yield one positive and one negative (unphysical, since $x>0$) solution. The roots \eqref{eq:app-x1} are real if and only if $2\sqrt{3}\,\xi^{2}-3\Omega \geq 0$, which is the upper bound in \eqref{eq:app-window}. Note that this reality condition constrains $\Omega$ only from above; indeed, nothing in the root structure singles out $\Omega=\xi^{2}$.

Setting $\Omega=\xi^{2}$ in \eqref{eq:app-Vlargefield}, the potential becomes a perfect square,
\begin{equation}
  V(x)\big|_{\Omega=\xi^{2}}= \frac{9M_P^{4}}{4\alpha}\left(1-\frac{\xi x}{2}\right)^{2}\,,
  \label{eq:app-square}
\end{equation}
with a double zero at $x_{\rm min}=2/\xi$, i.e.~$\rho_{\rm min}=\sqrt{3/2}\;M_P\ln(2/\xi)$, which is the large-field limit
of the exact relation $\sinh^{2}u_{\rm min}=\xi/(2\Omega)$. In the same limit, the quartic factorises,
\begin{equation}
  \left(\xi x-2\right)^{2}\left[\,4\xi^{2}x^{2}-3\left(\xi x-2\right)^{2}\right] = 0 \,,
  \label{eq:app-factor}
\end{equation}
so $x=2/\xi$ is a double root and the remaining factor yields the two roots $x=\pm 2\sqrt{3}/\big[(2\pm\sqrt{3})\,\xi\big]$. The double root is spurious as, at $\Omega=\xi^{2}$, the slow-roll parameter \eqref{eq:app-eps} reduces to
\begin{equation}
  \epsilon_{V}(x)\big|_{\Omega=\xi^{2}}
  = \frac{4\,\xi^{2}x^{2}}{3\left(\xi x-2\right)^{2}}
  \;\xrightarrow[\;x\to 2/\xi\;]{}\;\infty \,,
  \label{eq:app-pole}
\end{equation}
i.e.~\ $x=2/\xi$ is a pole of $\epsilon_{V}$ (namely, the minimum of the potential). Of the remaining roots, $x=-(4\sqrt{3}+6)/\xi$ is negative, so the unique physical solution is
\begin{equation}
  x_{\rm end} =\frac{4\sqrt{3}-6}{\xi} \;\approx\; \frac{0.928}{\xi}\,.
  \label{eq:app-xend}
\end{equation}
One may check that \eqref{eq:app-xend} coincides with the
$\Omega\to\xi^{2}$ limit of $x_{1}^{(-)}$ in \eqref{eq:app-x1}.

Since $x_{\rm end}<x_{\rm min}$ for any $\xi>0$, inflation ends regularly. The field exits slow roll at $x_{\rm end}$, well before reaching the minimum, and subsequently oscillates about $\rho_{\rm min}$. For the benchmark $\xi=10^{-2}$, we find
\begin{equation}
  \rho_{\rm end}=\sqrt{\tfrac{3}{2}}\,M_P\ln(92.8)\approx 5.55\,M_P \,, \quad
  \rho_{\rm min}=\sqrt{\tfrac{3}{2}}\,M_P\ln(200)\approx 6.49\,M_P \,,
  \label{eq:app-numbers}
\end{equation}
in agreement with the numerical integration, which gives $\rho_{\rm end}\approx 5.56\,M_P < \rho_{\rm min}\approx 6.50\,M_P$. The boundary $\Omega=\xi^{2}$ is therefore a regular part of the physically viable region as far as the inflationary dynamics is concerned.

\bibliographystyle{JHEP}
\bibliography{biblio}
\end{document}